\pdfoutput=1
\documentclass[rmp,aps,amsmath,amssymb,nofootinbib]{revtex4-2}
\usepackage[english]{babel}
\usepackage{graphicx,amsmath,relsize,epstopdf,color,mathtools,bm,newtxtext,newtxmath,rotating,dsfont,stackrel}
\usepackage[hyphenbreaks]{breakurl}
\usepackage[colorlinks=true,linkcolor=blue,citecolor=blue,urlcolor =blue]{hyperref}
\usepackage{hyperref}
\usepackage{relsize}
\usepackage{setspace}

\usepackage{lineno}

\newcommand{\eq}[1]{\begin{equation}\begin{aligned}#1\end{aligned}\end{equation}}
\newcommand{\iu}{\text{i}}
\newcommand{\eu}{\text{e}}
\newcommand{\ha}{\hat{a}}
\newcommand{\had}{\hat{a}^\dagger\vphantom{a}}
\newcommand{\hb}{\hat{b}}
\newcommand{\hbd}{\hat{b}^\dagger\vphantom{a}}

\newcommand{\ket}[1]{\lvert#1\rangle}

\newcommand{\bra}[1]{\langle#1\rvert}

\newcommand{\braket}[2]{\langle#1\rvert#2\rangle}
\newcommand{\expct}[1]{\langle#1\rangle}
\newcommand{\Tr}{\mathop{\mathrm{Tr}} \nolimits}

\newcommand{\Var}{\mathop{\mathrm{Var}}\nolimits}
\newcommand{\Cov}{\mathop{\mathrm{Cov}} \nolimits}
\newcommand{\vac}{\ket{\mathrm{vac}}}
\newcommand{\btheta}{\pmb{\theta}}

\newcommand{\RE}{\mathop{\mathrm{Re}} \nolimits}
\newcommand{\IM}{\mathop{\mathrm{Im}} \nolimits}

\begin{document}
	
	\title{Quantum Polarimetry}
	\author{Aaron Z. Goldberg}
	\affiliation{National Research Council of Canada, 100 Sussex Drive, Ottawa, Ontario K1A 0R6, Canada}

	\begin{abstract}
		Polarization is one of light's most versatile degrees of freedom for both classical and quantum applications. The ability to measure light's state of polarization and changes therein is thus essential; this is the science of polarimetry. It has become ever more apparent in recent years that the quantum nature of light's polarization properties is crucial, from explaining experiments with single or few photons to understanding the implications of quantum theory on classical polarization properties. We present a self-contained overview of quantum polarimetry, including discussions of classical and quantum polarization, their transformations, and measurements thereof. We use this platform to elucidate key concepts that are neglected when polarization and polarimetry are considered only from classical perspectives.
	\end{abstract}
	
	\maketitle


\tableofcontents{}
\section{Introduction}
Polarization, in its essence, describes the behaviour of the electric field within an electromagnetic wave. This behaviour can readily be manipulated \cite{ChoquetteLeibenguth1994,Grier2003,Straufetal2007} and measured \cite{Hauge1976,Mishchenkoetal2011}, making it useful for encoding classical and quantum information. Polarization has successfully been used for communication and metrology
and is so ubiquitous that it is treated in a large number of authoritative works \cite{Wiener1930,Wolf1959,McMaster1961,ClarkeGrainger1971,AzzamBashara1977,Collett1992,MandelWolf1995,Huard1997,Brosseau1998,Jackson1999,Goldstein2003,Collett2005,BrosseauDogariu2006,Gil2007,Brosseau2010,Brown2011,KumarGhatak2011,KligerLewis2012,Pye2015,GilOssikovski2016}.

Because light's polarization changes when it interacts with an object, be it through reflection or transmission, it is readily used for characterizing objects without disturbing the latter. Polarimetry and its relative ellipsometry \cite{Azzam2011} do just that: they characterize substances by the unique changes they impart on light's polarization degrees of freedom \cite{Collett1992}. Highly precise measurements of polarization and its changes, i.e., polarimetry, have found applications in photonics \cite{Yoonetal2020}, bioimaging \cite{Dulketal1994,Ghoshetal2009,Tuchin2016,FirdousAnwar2016}, oceanography \cite{VossFry1984}, remote sensing \cite{Tyoetal2006}, astronomy \cite{Tinbergen2005,Dulketal1994}, and beyond.

It is now taken for granted by practitioners of quantum optics that light is fundamentally described by a quantum field \cite{Glauber1963,Glauber1963quantumtheorycoherence,Sudarshan1963}. Light's polarization properties are no different, also fundamentally arising from quantum theory \cite{Fano1949,FalkoffMacDonald1951,Fano1954,JauchRohrlich1955,McMaster1961,Collett1970}, and have been the subject of recent reviews \cite{Karassiov2007review,Luis2016,Goldbergetal2021polarization}, where the nuances of quantum polarization are seen to overthrow certain concepts from classical polarization \cite{Klyshko1992,Usachevetal2001,delaHozeal2014}. In fact,  quantum states may appear to be ``classically unpolarized'' while possessing “hidden” polarization properties, leading to
quantum polarization effects with no classical parallel \cite{PrakashChandra1971,Klyshko1992,Klyshko1997,Tsegayeetal2000,Usachevetal2001,Bushevetal2001,Luis2002,AgarwalChaturvedi2003,Bjorketal2015,Bjorketal2015PRA,LuisDonoso2016,ShabbirBjork2016,GoldbergJames2017,Bouchardetal2017,GoldbergJames2018Euler,Goldbergetal2020extremal}.
Quantum polarization has been used for quantum key distribution \cite{Bennettetal1992,Mulleretal1993}, Einstein-Podolsky-Rosen tests \cite{Kwiatetal1995}, quantum teleportation \cite{Bouwmeesteretal1997}, quantum tomography
\cite{Jamesetal2001}, weak value amplification \cite{Hallajietal2017}, and more.
However, the study of the \textit{changes} in quantum polarization have not, to our knowledge, been the focus of any major review, which is a void that must especially be filled due to the proliferation of recent experiments on polarimetry with explicitly quantum mechanical states of light \cite{Mitchelletal2004,Bogdanovetal2004,Toussaintetal2004,Grahametal2006,Ozaetal2010,Altepeteretal2011,Slussarenkoetal2017,Dayanooshetal2018,Yoonetal2020,Rosskopfetal2020,Sunetal2020arxiv,Zhangetal2021arxiv}. We set forth to present a complete picture of quantum polarimetry in this work.

There are a number of steps required for understanding quantum polarimetry. First, we briefly review light's polarization properties from a classical standpoint in Section \ref{sec:classical polarization}, making note of the different geometrical and mathematical representations of these properties and various schemes for determining them in the laboratory. Next, we discuss the mathematical structures describing classical changes in polarization, which lead to discussions of the physically viable and forbidden transformations. These structures are then directly useful for characterizing materials through polarimetry. In turn, we introduce light's polarization properties from a quantum mechanical perspective in Section \ref{sec:quantum polarization}, explaining the correspondences with classical polarization properties and giving examples of the peculiar properties about which classical polarization is ignorant. The stage is now set for quantum polarimetry, beginning with a rigorous discussion of the possible quantum mechanical transformations underlying classical polarimetry. Additionally, quantum polarimetry can then incorporate the tools of quantum estimation theory, so we discuss in Section \ref{sec:quantum polarimetry estimation} the possible quantum enhancements in polarimetry. We also include discussions of techniques for measuring quantum polarization properties and how they compare to classical techniques.
Deviations from and fortuitous agreement with classical polarization predictions abound, but they tend to be neglected in treatments of classical polarization, so we review them in Section \ref{sec:classical intuitions} and stress some nuances that are often overlooked.
We hope this reference will serve as a useful guide in the rapidly evolving world of polarization.

\section{Classical Polarization}
\label{sec:classical polarization}
Light's polarization degrees of freedom stem from Maxwell's equations. The simplest electromagnetic field obeying these equations is a plane wave:
\eq{
    \mathbf{E}(\mathbf{r},t)=\mathcal{E}_0\left(a\mathbf{e}_a+b\mathbf{e}_b\right)\eu^{\iu\left(\mathbf{k}\cdot\mathbf{r}-\omega t\right)}.
    \label{eq:plane wave}
} Here, the plane wave is travelling in direction $\mathbf{k}$ within some large quantization volume $V$, $\mathbf{e}_a$ and $\mathbf{e}_b$ are mutually orthogonal unit vectors that are orthogonal to $\mathbf{k}$, $\omega$ is the wave's angular frequency, all dimensionful constants are absorbed into $\mathcal{E}_0=\sqrt{\hbar\omega/2V\varepsilon_0}$, and $\varepsilon_0$ is the permittivity of free space. The quantity $\mathbf{E}(\mathbf{r},t)$ represents the analytic signal of the electric field and the true physical quantity that can be measured is the real part thereof \cite{BornWolf1999}; the positive-frequency component of the true field is given by $\mathbf{E}(\mathbf{r},t)/2$ \cite{Glauber1963quantumtheorycoherence}. The polarization properties of the electromagnetic field correspond to the path taken by the tip of the electric field vector $\mathbf{E}(\mathbf{r},t)$ over time in the $\mathbf{e}_a$-$\mathbf{e}_b$ plane.

Electromagnetic fields carry energy in direction $\mathbf{k}$, which we set to be the $\mathbf{z}$ direction for convenience. The electric field for a plane wave then traverses an ellipse in the $x$-$y$ plane. This ellipse is parametrized by the measured $a$ and $b$ components of the electric field, $E_a=\RE\left[\mathbf{e}_a\cdot \mathbf{E}(\mathbf{r},t)\right]$ and $E_b=\RE\left[\mathbf{e}_b\cdot \mathbf{E}(\mathbf{r},t)\right]$, through \cite{BornWolf1999}
\eq{
    \left(\frac{E_a}{\left|a\right|}\right)^2+\left(\frac{E_b}{\left|b\right|}\right)^2-2\frac{E_a E_b}{\left|ab\right|}\cos\delta=\mathcal{E}_0^2\sin^2\delta.
} We observe that the relative phase $\delta = \arg(a/b)$ and the amplitudes $|a|$ and $|b|$ completely describe the orientation
and relative sizes of the axes of the ellipse, while $\mathcal{E}_0$ governs the size of the ellipse. This equation is explicitly independent of time, as it simply governs the shape of the ellipse, and the electric field itself rotates around the ellipse over time when view along the direction of propagation. This is the sense in which the electric field's tip traverses the $\mathbf{e}_a$-$\mathbf{e}_b$ plane over time; we do not have to worry about defining complex amplitudes therein.

The polarization ellipse is commonly defined using the $x$ and $y$
components of $\mathbf{E}$ for $a$ and $b$, so as to categorize beams of light by the ellipticity of their polarization ellipses. We will
adopt the slightly nonstandard notation in which $\mathbf{e}_a$ and $\mathbf{e}_b$ refer to right- and left-handed circularly
polarized light, respectively, through the unit vectors $\mathbf{e}_a=\mathbf{e}_{\mathrm{R}}=\left(\mathbf{x}-\iu \mathbf{y}\right)/\sqrt{2}$ and $\mathbf{e}_b=\mathbf{e}_{\mathrm{L}}=\left(\mathbf{x}+\iu \mathbf{y}\right)/\sqrt{2}$, so in
this notation the shape of the polarization ellipse itself does not follow the nomenclature describing the polarization
as ``linear'' or ``circular.'' One  reason for choosing circularly polarized light as our
fundamental elements is that these are the components that directly feature in the interactions mediated by beams of light to enact transitions
between atomic sublevels \cite{Steck2016}. We will elucidate other motivations for prioritizing the circular components in our later descriptions of polarization.

Polarization is easy to measure because it can be determined through intensity measurements alone. In terms of energy flow, the flux received in a given area averaged over a time that involves
many oscillations of $1/\omega$ is given by the \textit{irradiance} of the field
\eq{
    I\propto \mathbf{E}^*\cdot\mathbf{E}=\mathcal{E}_0^2\left(\left|a\right|^2+\left|b\right|^2\right),
    \label{eq:intensity irradiance}
} where $\,^*$ denotes the complex conjugate. The field's intensity, in slight contrast, is the energy flux per unit area in the $x$-$y$ plane (i.e., perpendicular to $\mathbf{k}$), but we will presently refer to $I$ as the intensity of the field, 
per common parlance [e.g., in the original work by \textcite{Stokes1852} and the authoritative work by \textcite{BornWolf1999}]; we can equivalently assert that all detectors measuring irradiance are in the $x$-$y$ plane. We can, further, define the intensities of the various components of the light through
\eq{
    I_i=\left|\mathbf{e}_i\cdot\mathbf{E}\right|^2.
}

\subsection{Characterizing Polarization}
Light's polarization properties can be described in many equivalent ways. For example, one may determine the total intensity $I$ as well as the relative phase and amplitude of the two components $a$ and $b$. Similarly, one may consider the orientation and ellipticity of the polarization ellipse through the combined parameter
\eq{
    \frac{a}{b}=\left|\frac{a}{b}\right|\eu^{\iu\delta}.
} Employing units that absorb the proportionality constant and $\mathcal{E}_0$ from Eq. \eqref{eq:intensity irradiance}, we can also describe light's polarization using the four Stokes parameters, which codify the intensity differences between three pairs of orthogonal components of the light:
\eq{
    S_0&=\frac{I_{\mathrm{H}}+I_{\mathrm{V}}}{2}=\frac{I_{\mathrm{D}}+I_{\mathrm{A}}}{2}=\frac{I_{\mathrm{R}}+I_{\mathrm{L}}}{2}=\frac{\left|a\right|^2+\left|b\right|^2}{2} ,\\
    S_1&=\frac{I_{\mathrm{H}}-I_{\mathrm{V}}}{2}=\left|ab\right|\cos\delta ,\\
    S_2&=\frac{I_{\mathrm{D}}-I_{\mathrm{A}}}{2}=\left|ab\right|\sin\delta ,\\
    S_3&=\frac{I_{\mathrm{R}}-I_{\mathrm{L}}}{2}=\frac{\left|a\right|^2-\left|b\right|^2}{2}.
    \label{eq:Stokes classical}
} We have now made use of the horizontal and vertical unit vectors $\mathbf{e}_{\mathrm{H}}=\mathbf{x}$ and $\mathbf{e}_{\mathrm{V}}=\mathbf{y}$ and their diagonal and antidiagonal counterparts $\mathbf{e}_{\mathrm{D}}=\left(\mathbf{x}+\mathbf{y}\right)/\sqrt{2}$ and $\mathbf{e}_{\mathrm{A}}=\left(\mathbf{x}-\mathbf{y}\right)/\sqrt{2}$. 
Other definitions of the Stokes parameters differ from ours by a nonessential factor of 2.
We note that, even though Stokes
first published work on these parameters in 1852 \cite{Stokes1852}, they were only rediscovered by Soleillet in 1929 \cite{Soleillet1929} and made famous by Chandrasekhar in the
1940s \cite{Chandrasekhar1950}. In the interim, the so-called Poincar\'e sphere was developed, which describes plane waves' polarization by a point on the unit sphere \cite{Poincare1889}. This is made apparent by realizing that the Stokes parameters for a plane wave obey
\eq{
    S_0^2=S_1^2+S_2^2+S_3^2,
    \label{eq:Stokes sphere}
} so the polarization properties other than the total intensity are completely specified by the two angular coordinates of the polarization vector $\mathbf{S}=\left(S_1,S_2,S_3\right)^\top$, where $\,^\top$ denotes the transpose. 
As well, the Jones vector formalism was also developed, expressing the electric field in the circular basis with the time and space dependence removed, or expressed in a frame counter-rotating with angular frequency $\left(\mathbf{k}\cdot\mathbf{r}-\omega t\right)$, as \cite{Jones1941}
\eq{
    \mathbf{A}=\mathbf{E}\eu^{-\iu\left(\mathbf{k}\cdot\mathbf{r}-\omega t\right)}=\begin{pmatrix}
        a\\b
    \end{pmatrix}.
    \label{eq:quantized Jones vector}
}
Soon after, the coherency matrix description of polarization was developed \cite{Wolf1959}:
\eq{
    \boldsymbol{\Psi}=\mathbf{A}\mathbf{A}^\dagger=\begin{pmatrix}a^* a&b^*a\\a^*b&b^*b
    \end{pmatrix}.
} where $\,^\dagger$ denotes the Hermitian conjugate. The Stokes parameters are related to the coherency matrix through the compact expression
\eq{
    S_\mu=\frac{1}{2}\mathbf{A}^\dagger \sigma_\mu\mathbf{A}=\frac{1}{2}\Tr\left(\boldsymbol{\Psi}\sigma_\mu\right),
} using the Pauli matrices $\sigma_\mu$ and letting Greek indices run from $0$ to $3$. This description in terms of the Pauli matrices is another reason for preferentially choosing the circular basis in our description of the electric field. We will see in our discussions of polarization from a quantum perspective the mathematical usefulness of the physically irrelevant factors of $2$ that we have incorporated into all of our expressions.

Far from any source, plane waves are a good approximation to any freely propagating monochromatic wave. 
The polarization formalism developed for planes waves, fortunately, extends to quasimonochromatic light. Quasimonochromatic waves with average frequencies $\bar{\omega}$ are now characterized by their two complex components at a particular position $z$ (again, in units of $\mathcal{E}_0$)
\eq{
E_{a}(t)=a(t)
\eu^{-\iu \bar{\omega} t}\quad \mathrm{and}\quad E_{b}(t)=b(t)
\eu^{-\iu \bar{\omega} t}.
} Here, in contradistinction to Eq. \eqref{eq:plane wave}, the amplitudes $a(t)$ and $b(t)$ depend on time, but they only vary on timescales that are long compared to $1/\bar{\omega}$. We can then define the coherency matrix and Stokes parameters as before by taking a time average that includes many oscillations of $\bar{\omega}$, which, under the standard assumptions of stationarity and ergodicity, is equivalent to taking an ensemble average $\langle\cdot\rangle$:
\eq{
\boldsymbol{\Psi}=\expct{\mathbf{A}\mathbf{A}^\dagger}=
\begin{pmatrix}
	\expct{a^* a } & \expct{b^*a } \\ \expct{a^* b} & \expct{b^*b }
\end{pmatrix}
} and
\eq{
	S_\mu= \frac{1}{2}\Tr\left(\boldsymbol{\Psi} \sigma_\mu\right) = \frac{1}{2}\expct{\mathbf{A}^\dagger\sigma_\mu\mathbf{A} }.
	\label{eq:Stokes ensemble definition}
} Now, the Stokes vectors no longer satisfy Eq. \eqref{eq:Stokes sphere} and are, thus, not constrained to the surface of the Poincar\'e sphere. Instead, the Stokes parameters define a vector pointing somewhere inside the Poincar\'e sphere of radius $S_0$. The orientation of the vector $\mathbf{S}$ has two angular coordinates  as before. Now, the length of $\mathbf{S}$ defines a new parameter called the degree of polarization \cite{Wiener1930,BillingsLand1948,Walker1954,Wolf1954,Wolf1959}
\eq{
p= \frac{\sqrt{\mathbf{S}\cdot\mathbf{S}}}{S_0} .
} The degree of polarization ranges from $0$, for completely unpolarized light, to $1$, for perfectly polarized light. This can be inferred from the positivity of the coherency matrix, which is equivalent to the constraint
\eq{
 S_1^2+S_2^2+S_3^2 \leq S_0^2.
\label{eq:stokes squared inequality}
}
When the degree of polarization is less than $1$, the electric field no longer traces out a perfect ellipse over time, with decreasing $p$ implying an increasingly erratic behaviour for this vector.
We will see later that these properties are important to revisit with the quantum theory of polarization.

\subsubsection{Polarized States}
We saw before that plane waves have $p=1$, making them perfectly polarized. Any other beam of light with $p=1$ is again perfectly polarized, which can be determined in a number of different ways.

To start, a beam of light whose electric field traces a closed ellipse over time is completely polarized. Next, a beam whose Stokes vector lies on the surface of the Poincar\'e sphere is fully polarized. This means that the polarization vector must obey $\mathbf{S}=S_0\mathbf{e}$ for some unit vector $\mathbf{e}$, such that, defining the ``vector'' of Stokes parameters $\mathcal{S}=\left(S_0,S_1,S_2,S_3\right)^\top$,
\eq{
    \mathcal{S}_{\mathrm{pol}}=S_0\begin{pmatrix}
        1\\
        \sin\Theta\cos\Phi\\\sin\Theta\sin\Phi\\\cos\Theta
    \end{pmatrix}.
    \label{eq:stokes pol}
} Here, the parameters $\Theta$ and $\Phi$ correspond to the angular coordinates of $\mathbf{S}$ and thus to the direction of the Stokes vector on the Poincar\'e sphere.

In terms of the coherency matrix, polarized light must be a rank-one projector. This is because any single Jones vector $\mathbf{A}$ corresponds to completely polarized light and only through an ensemble average of nonparallel Jones vectors does the coherency matrix lose polarization. As such, perfectly polarized light has the determinant of $\boldsymbol{\Psi}$ vanish, through
\eq{
    \boldsymbol{\Psi}_{\mathrm{pol}}=2S_0\begin{pmatrix}
        \cos\frac{\Theta}{2}\\\sin\frac{\Theta}{2}\eu^{\iu \Phi}
    \end{pmatrix}\begin{pmatrix}
        \cos\frac{\Theta}{2}&\sin\frac{\Theta}{2}\eu^{-\iu \Phi}
    \end{pmatrix},
    \label{eq:coherency pol}
} where the angular coordinates $\Theta$ and $\Phi$ are the same as for the Stokes parameters.

Translating between different states of light with perfect polarization is equivalent to changing the angular coordinates $\Theta$ and $\Phi$. Mathematically, this corresponds to rotating the polarization vector $\mathbf{S}$ and the coherency matrix $\boldsymbol{\Psi}$ or to changing the polarization ellipse, while these translations can be physically enacted by common devices such as waveplates.

\subsubsection{Unpolarized States}
Unpolarized light has $p=0$ and is, in some sense, the most different from plane waves. The trace of its electric field does not return to its starting position at the same angle, instead exhibiting chaotic behaviour to cover the entire disk. As with perfectly polarized light, unpolarized light can be determined in a variety of manners.

The most important property of unpolarized light is that it is unchanged by physical devices that rotate the direction of polarization. For this to be true of a Stokes vector, it must have a vanishing polarization vector $\mathbf{S}=\mathbf{0}$:
\eq{
    \mathcal{S}_{\mathrm{unpol}}=S_0\begin{pmatrix}
        1\\
        0\\0\\0
    \end{pmatrix}.
    \label{eq:Stokes unpol}
} Similarly, the coherency matrix must be unchanged by rotations, meaning that it must be proportional to the identity matrix:
\eq{
    \boldsymbol{\Psi}_{\mathrm{unpol}}=S_0\begin{pmatrix}
        1&0\\0&1
    \end{pmatrix}.
    \label{eq:coherency unpol}
} The angular coordinates $\Theta$ and $\Phi$ are undefined for such light and no waveplate can affect the polarization properties of unpolarized light.

\subsubsection{Decomposition of Light into Polarized and Unpolarized Elements}
The classical polarization properties of quasimonochromatic light are completely described by four parameters, which are often organized into
the polarization direction information, given by the direction of $\mathbf{S}$, the degree of polarization, given by $p$, and the total intensity, given by $S_0$. There is another useful interpretation of these parameters that follows from the linearity properties of independent electromagnetic waves. Superposing more than one wave leads to the summing of their electric field amplitudes. Then, ensemble averages over components from different fields all vanish when those fields are independent. For example, if the first field has amplitudes $a^{(1)}$ and $b^{(1)}$ and the second $a^{(2)}$ and $b^{(2)}$, independence of the two fields dictates that
\eq{
\expct{a^{(1)*} a^{(2)}}=\expct{a^{(1)*}}\expct{ a^{(2)}}=0=\expct{a^{(1)*}}\expct{ b^{(2)}}=\expct{a^{(1)*} b^{(2)}}.
} This directly implies that the polarization properties of the superposed fields are equal to the sums of their respective components from the two contributing fields: the Stokes parameters are comprised by \eq{ S_\mu^{(1+2)}=S_\mu^{(1)}+S_\mu^{(2)} 
} and
the coherency matrix takes the form
\eq{
\boldsymbol{\Psi}^{(1+2)}=\boldsymbol{\Psi}^{(1)}+\boldsymbol{\Psi}^{(2)}.}
This lets us imagine every partially polarized field ($0<p<1$) to have arisen from the \textit{incoherent} superposition of a field that is completely unpolarized with a field that is perfectly polarized, with the relative weight in the sum being given by the degree of polarization $p$. 

How is this constructed? We take convex combinations of the polarized and unpolarized components from Eqs. \eqref{eq:stokes pol}, \eqref{eq:coherency pol}, \eqref{eq:Stokes unpol}, and \eqref{eq:coherency unpol}. In terms of Stokes parameters, we find the general expression
\eq{
    \mathcal{S}=p\mathcal{S}_{\mathrm{pol}}+\left(1-p\right)\mathcal{S}_{\mathrm{unpol}},
    \label{eq:Stokes decomposition}
} and we find a similar decomposition in terms of the coherency matrix
\eq{
    \boldsymbol{\Psi}=p\boldsymbol{\Psi}_{\mathrm{pol}}+\left(1-p\right)\boldsymbol{\Psi}_{\mathrm{unpol}}.
    \label{eq:coherency decomposition}
} This implies that, regardless of the physical origin of a beam of light, we can always consider it to have arisen from the probabilistic mixture of a pefectly polarized and a completely unpolarized beam of light, with the probability of the former being given by the degree of polarization. We stress here that this is a \textit{local} description, in that the relative weight in the probabilistic mixture changes upon propagation \cite{James1994} and across the $x$-$y$ plane \cite{Korotkovaetal2008}, so we cannot, in general, consider the polarized and unpolarized fractions to each simply propagate as independent beams of light for arbitrary beam profiles \cite{Korotkovaetal2008}. The locality caveat can be removed if one looks at the polarization properties integrated over the $x$-$y$ plane, as those do not change upon propagation \cite{James1994}.

In all of the decompositions, there is one parameter governing the total intensity, two parameters governing
the orientation of the perfectly polarized component, and a final parameter, the degree of polarization, responsible
for the relative contributions of the polarized and unpolarized components. These degrees of freedom can be
used to specify a wide range of quantum states underlying the same classical beams of light. Extra degrees of
freedom beyond the four in the decompositions of Eqs. \eqref{eq:Stokes decomposition} and \eqref{eq:coherency decomposition} give rise to numerous novel polarization
phenomena that can only be investigated quantum mechanically, a topic to which we will give much consideration.

\subsection{Measuring Polarization}
\begin{figure*}
    \centering
    \includegraphics[width=\textwidth]{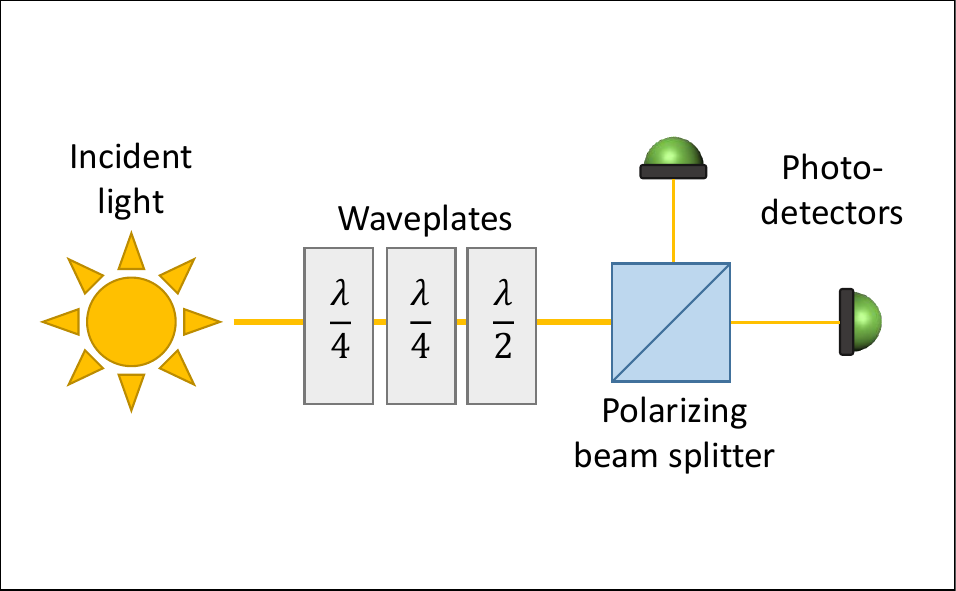}
    \caption{SU(2) gadget for measuring Stokes parameters. The polarizing beam splitter (PBS) separates the horizontal and vertical components of the light to have their intensities measured by separate photodetectors. The two quarter waveplates (``$\lambda/4$'') and one half waveplate (``$\lambda/2$'') are sufficient to enact any polarization rotation, such that the PBS can split the light into any pair of orthogonal polarization components. The original scheme by \textcite{SimonMukunda1990} used two quarter waveplates and two half waveplates but modern schemes reduce this number \cite{Schillingetal2010} and the order of the waveplates is immaterial \cite{SimonMukunda1990}.}
    \label{fig:SU2 gadget}
\end{figure*}

Determining the four Stokes parameters is tantamount to providing complete knowledge of a classical polarization state. As per Eq. \eqref{eq:Stokes classical}, these parameters can be found by measuring six different intensities and computing various differences and sums between them. Fortunately, polarizing beam splitters (PBSs) can be used to spatially separate two orthogonal polarization components of a beam of light to each have their intensities measured by a photodector, while waveplates can be used to rotate a beam's polarization to properly arrange the two components of a beam to be measured. This led to the universal SU(2) gadget for measuring polarization the Stokes parameters \cite{SimonMukunda1990}, depicted in Fig. \ref{fig:SU2 gadget}.

The SU(2) gadget is useful for serially measuring all of the Stokes parameters, but what if one desires to measure all four parameters simultaneously? One trick is to first divide the light into three beams and then measure three different orthogonal pairs of polarization intensities as with a single SU(2) gadget \cite{Azzam1982,Azzam1985, AzzamDe2003}. Alternatively, one can employ an SU(2) gadget whose waveplates rotate continuously, such that one can recover the Stokes parameters from the ensuing interferograms \cite{Goldstein1992,Schaeferetal2007}. More intricacies are required if one wants to measure the quantum polarization properties of a beam of light, such as by appropriately splitting the light into eight beams \cite{AlodjantsArakelian1999,Alodjantsetal1999}.

Instead of waveplates, liquid crystals can also be used in creating an SU(2) gadget. Then, instead of rotating waveplates to change the polarization direction, one can adjust the voltage applied to the liquid crystals to enact the same transformation \cite{Luetal2015}. This method is a promising new tool in the polarimeter's arsenal \cite{SuWang2021}.

According to classical electrodymanics, all of the polarization components can be measured with arbitrary precision. This is in contradistinction with quantum polarization, where we will see explicitly that there exists a lower limit to the precision with which all of the Stokes parameters can be simultaneously measured.

\subsection{Changes in Polarization}
Polarimetry is useful because changes in polarization contain useful information about objects being measured. There are a few related methods for arranging this information, which we explore in turn.
\subsubsection{Jones Matrix Calculus}
The most basic question to ask is how an electric field changes when it interacts with an object. When these transformations are linear, the most general possible result is
\eq{
    \mathbf{A}^{(\mathrm{in})}\to \mathbf{A}^{(\mathrm{out})}=\mathbf{J}\mathbf{A}^{(\mathrm{in})}\qquad \iff \qquad \mathbf{E}^{(\mathrm{in})}\to \mathbf{E}^{(\mathrm{out})}=\mathbf{J}\mathbf{E}^{(\mathrm{in})},
} where $\mathbf{J}$ is a $2\times 2$ complex matrix known as the Jones matrix. These transformations are deemed ``nondepolarizing'' polarization transformations because they do not change the degree of polarization of perfectly polarized states. A range of physical and linguistic arguments in favour of adopting the nomenclature ``deterministic'' instead of ``nondepolarizing'' are due to \textcite{Simon1990}, but we will see that this is at odds with the standard quantum definition of deterministic transformations.

When the electric field undergoes a Jones matrix transformation, the coherency matrix transforms as
\eq{
    \boldsymbol{\Psi}^{(\mathrm{in})}\to \boldsymbol{\Psi}^{(\mathrm{out})}=\mathbf{J}\boldsymbol{\Psi}^{(\mathrm{in})}\mathbf{J}^\dagger.
    \label{eq:coherency pure jones}
} This equation now holds even for quasimonochromatic incident light. Then, there are some linear transformations that cannot be described by a single Jones matrix transformation. Instead, the most general linear transformation of the coherency matrix is
\eq{
    \boldsymbol{\Psi}^{(\mathrm{in})}\to \boldsymbol{\Psi}^{(\mathrm{out})}=\sum_i \mathbf{J}_i\boldsymbol{\Psi}^{(\mathrm{in})}\mathbf{J}_i^\dagger,
    \label{eq:coherency multiple Jones}
} where each $2\times 2$ complex matrix $\mathbf{J}_i$ describes a deterministic transformation. These new transformations with more than a single $\mathbf{J}_i$ are termed ``depolarizing'' or ``nondeterministic'' and distinctions between the two types of transformations have been investigated thoroughly \cite{Kimetal1987,Simon1990,KuscerRibaric1959,AbhyankarFymat1969,FryKattawar1981,Barakat1981,Simon1982,GilBernabeu1985,Simon1987,Cloude1990,BrosseauBarakat1991,Kostinski1992,vanderMee1993,Brosseauetal1993,Kostinskietal1993,AndersonBarakat1994,Hovenier1994,CloudePottier1995,Raoetal1998I,Raoetal1998II,Gil2000,EspinosaLuna2007,Zakerietal2013}.

What cannot be described by convex combinations of Jones matrix transformations are linear transformations of the form
\eq{
    \boldsymbol{\Psi}^{(\mathrm{in})}\to \boldsymbol{\Psi}^{(\mathrm{out})}=\sum_i \lambda_i \mathbf{J}_i\boldsymbol{\Psi}^{(\mathrm{in})}\mathbf{J}_i^\dagger
    \label{eq:coherency multiple Jones with weights}
} with negative ``weights'' $\lambda_i$. The standard assumption is that only transformations with positive weights $\lambda_i>0$ are physically viable \cite{Cloude1986}, which is equivalent to requiring the transformations of the coherency matrix to be completely positive in the sense of Choi's theorem \cite{AhnertPayne2005,Aielloetal2007,Sudhaetal2008,GamelJames2011}. This allows us to interpret nondeterministic transformations as probabilistic mixtures of Jones matrix transformations.

Transformations governed by a single Jones matrix can result from a number of physical processes. Standard electrodynamic theory dictates that different polarizations of light have different probabilities of being reflected off of a surface and, in terms of transmission, different polarization components experience different indices of refraction, leading to the development of a relative phase between the components \cite{Jackson1999}. 
In turn, different physical interactions lead to different transformations of the coherency matrix. By preparing a variety of states of light with known input coherency matrices and measuring the output coherency matrices after the physical interaction, one can identify the full Jones matrix description of the interaction and thereby classify the object with which the light interacted.

\subsubsection{Mueller Matrix Calculus}
Linear polarization transformations are naturally described by the Stokes vector transformation
\eq{
    \mathcal{S}^{(\mathrm{in})}\to \mathcal{S}^{(\mathrm{out})}=\mathbf{M}\mathcal{S}^{(\mathrm{in})}.
    \label{eq:Mueller definition Stokes}
} Here, the $4\times 4$ real Mueller matrix $\mathbf{M}$ encodes all of the possible classical information that can be garnered from a linear polarization transformation. As with coherency matrices,
by preparing a variety of input states of light with known Stokes parameters and measuring the output Stokes parameters after the physical interaction, one can identify the full Mueller matrix description of the interaction and thereby classify the object with which the light interacted \cite{Hauge1980}. For example, in the type of polarimetry alternatively known as transmission ellipsometry, one compares the
measured $\mathbf{M}$ to various models of Mueller matrices for a medium through which the light
has passed \cite{AzzamBashara1977}. Linearity lets one assume that all sets of input Stokes parameters
transform according to an identical Mueller matrix when transmitted through the same physical
system, such that all $16$ components of $\mathbf{M}$ can be estimated using the same set of probe beams without any \textit{a priori} knowledge of the object being measured.

Mueller matrices corresponding to Jones matrix transformations can be calculated using
\eq{
    M_{\mu \nu}=\frac{1}{2}\Tr\left(\sigma_\mu\mathbf{J} \sigma_\nu\mathbf{J}^\dagger\right).
    \label{eq:Mueller from pure Jones}
} This compact correspondence is yet another motivation for singling out circular polarization in our basis elements. Similarly, Mueller matrices corresponding to nondeterministic polarization transformations can be calculated through
\eq{
    M_{\mu \nu}=\sum_i \lambda_i\frac{1}{2}\Tr\left(\sigma_\mu\mathbf{J}_i \sigma_\nu\mathbf{J}_i^\dagger\right).
    \label{eq:Mueller from Jones with weights}
} While it is clear that Mueller matrices can exist with $\lambda_i<0$, such as the matrix 
\eq{
    \mathbf{M}=\begin{pmatrix}
        1&0&0&0\\0&1&0&0\\0&0&1&0\\0&0&0&-\frac{1}{3}
    \end{pmatrix},
    \label{eq:Mueller reflection}
} which corresponds to a transformation of the coherency matrix with a negative weight
\eq{
    \boldsymbol{\Psi}^{(\mathrm{out})}=\frac{2}{3}\sigma_0\boldsymbol{\Psi}^{(\mathrm{in})}\sigma_0+\frac{1}{3}\sigma_1\boldsymbol{\Psi}^{(\mathrm{in})}\sigma_1+\frac{1}{3}\sigma_2\boldsymbol{\Psi}^{(\mathrm{in})}\sigma_2-\frac{1}{3}\sigma_3\boldsymbol{\Psi}^{(\mathrm{in})}\sigma_3,
} it is contended that such transformations are not physically viable. Evidence to the contrary \cite{Ossikovskietal2008} must be investigated on a case-by-case basis.

\subsubsection{Deterministic Transformations in terms of Scale Factor and SL$(2,\mathds{C})$}
There are four free complex components of a generic Jones matrix. However, the global phase of $\mathbf{J}$ is irrelevant in its action on the coherency matrix, as it cancels in Eq. \eqref{eq:coherency pure jones}, and does not show up in Mueller matrices, as it cancels in Eq. \eqref{eq:Mueller from pure Jones}, so deterministic polarizations transformations are determined by seven parameters. In this and the next subsection we present a number of equivalent ways to organize these parameters, each of which comes with its own physical insights.

Removing an overall multiplicative factor from a Jones matrix can always lead to the decomposition
\eq{
\mathbf{J}=t\pmb{J},
\label{eq:Jones scale factor}
} where $t\leq 1$ is an overall attenuation factor and $\pmb{J}$ belongs to the group of complex $2\times 2$ matrices with unit determinant SL$(2,\mathds{C})$, except in limiting situations in which $\det \mathbf{J}=0$. The factor $t$ does not affect the degree or direction of polarization, merely lowering $S_0$ by a factor of $t^2$ and shrinking the radius of the Poincar\'e sphere accordingly. One can separate measurement of the the scale factor from the other changes in polarization, by first measuring the total intensity of the transmitted beam, which transforms as
\eq{
S_0\to t^2 S_0 ,
} then separately determining the changes in orientation and length of the vector $\mathbf{S}$ normalized by the new intensity $t^2 S_0$ to determine the corresponding matrix $\pmb{J}$.

After dealing with the scale factor $t$, all Jones matrices are fundamentally related to the Lorentz group. This is because transformation matrices $\pmb{J}$ maintain the quadratic form \cite{Barakat1963}
\eq{
\mu^2 = S_0^2-\mathbf{S}\cdot\mathbf{S},
} as with four vectors in special relativity.
Deterministic polarization transformations can thus be though of as Lorentz transformations, with some transformations corresponding to rotations of the vector $\mathbf{S}$ and others corresponding to boosts along a particular axis. To rotate the polarization vector $\mathbf{S}$ by an angle $\Theta$ about the $\mathbf{n}$-axis, we employ a Jones matrix of the form
\eq{
\pmb{J}_{\mathrm{rot}}\left(\Theta,\mathbf{n}\right)=\exp\left(-\iu\Theta \mathbf{n}\cdot\boldsymbol{\sigma}/2\right).
\label{eq:Jones rotation}
} Equation \eqref{eq:Jones rotation} is an example of an SU(2) rotation generated by the vector of Pauli matrices $\boldsymbol{\sigma}=\left(\sigma_1,\sigma_2,\sigma_3\right)^\top$. 

When the electric field undergoes a rotation transformation $\mathbf{A}\to \pmb{J}_{\mathrm{rot}}\mathbf{A}$, its polarization properties correspondingly rotate in the Poincar\'e sphere. This is described by a Mueller matrix of the form
\eq{
\mathbf{M}_{\mathrm{rot}}\left(\Theta,\mathbf{n}\right)
=\begin{pmatrix}
    1&\mathbf{0}^\top\\
    \mathbf{0}&\mathbf{R}\left(\Theta,\mathbf{n}\right)
\end{pmatrix},
\label{eq:Mueller rotation}
} 
where $\mathbf{R}\left(\Theta,\mathbf{n}\right)$ is a standard $3\times 3$ rotation matrix that can be parametrized by a rotation angle $\Theta$ and axis of rotation $\mathbf{n}$, or equivalently by three Euler angles or another triad of parameters.
We explicitly give one parametrization of $\mathbf{R}$ through the Rodrigues rotation formula
\eq{
    R_{i j}\left(\Theta,\mathbf{n}\right)=\delta_{i j}\cos\Theta-\sum_{k=1}^3\epsilon_{i j k} n_k\sin\Theta+n_i n_j\left(1-\cos\Theta\right),
} where $\epsilon_{i j k}$ is the Levi-Civita tensor.
Rotations leaves the $S_0$ component unchanged, which accords with waveplates not changing the intensity of an incident beam of light, and maintains the length of $\mathbf{S}$, as such transformations do not affect light's degree of polarization. These properties carry through to the quantum description of polarization.
We remark that these rotations are polarization rotations, in the sense that they rotate the  polarization vector $\mathbf{S}$ in the Poincar\'e sphere and not in three-dimensional space; the effect of rotating a beam of light in three dimensions can be found in various works by~\textcite{BialynickiBirula1996,BialynickiBirulaBialynickiBirula2020}.

Similarly to the rotations expressed by Eq. \eqref{eq:Jones rotation}, to boost the Stokes vector along the $\mathbf{n}$-axis by a rapidity $\eta$, we employ the Jones matrix
\eq{
	\pmb{J}_{\mathrm{boost}}\left(\eta,\mathbf{n}\right)=\exp\left(\eta \mathbf{n}\cdot\boldsymbol{\sigma}/2\right) .
	\label{eq:Jones boost}
} Eq. \eqref{eq:Jones boost} differs from Eq. \eqref{eq:Jones rotation} by the crucial difference that the argument of the exponential is now real, so the Lorentz boost transformations can be thought of as rotations by an imaginary phase (or as \textit{hyperbolic} rotations). 

Readers versed in special relativity may be more familiar with the $4\times 4$ representation of the Lorentz group, which is directly provided by Mueller matrices. Using Eq. \eqref{eq:Mueller from pure Jones}, we can immediately match the various types of transformations with the more well-known representation. For example, a boost by ``rapidity'' $\eta$ along the $\mathbf{e}_3$ axis\footnote{We refer to three unit vectors by $\mathbf{e}_1=\left(1,0,0\right)^\top$, $\mathbf{e}_2=\left(0,1,0\right)^\top$, and $\mathbf{e}_3=\left(0,0,1\right)^\top$ so as to distinguish between polarization rotations on the Poincar\'e sphere and rotations in physical space spanned by $\mathbf{x}$, $\mathbf{y}$, and $\mathbf{z}$.} corresponds to the Mueller matrix
\eq{
\mathbf{M}_{\mathrm{boost}}\left(\eta,\mathbf{e}_3\right)=\exp\left[\eta \begin{pmatrix}
0 &0&0&1\\
0&0&0&0\\
0&0&0&0\\
1&0&0&0
\end{pmatrix}\right]=\begin{pmatrix}
	\cosh\eta &0&0&\sinh\eta\\
	0&1&0&0\\
	0&0&1&0\\
	\sinh\eta&0&0&\cosh\eta
\end{pmatrix},
} which is exactly the transformation matrix for ``boosts'' between reference frames with constant relative velocity.
Technically, 
Jones matrices correspond to the ``proper'' Lorentz group SO$\vphantom{a}(1,3)$, which removes the possibility of ``spatial'' reflections of the polarization vector $\mathbf{S}$ such as in Eq. \eqref{eq:Mueller reflection}. 
By disallowing spatial reflections, the proper Lorentz group is comprised only from rotations and boosts, each with three real parameters corresponding to the axis and strength of the transformations. This lets us arrange the seven free parameters of deterministic polarization transformations into:
\begin{itemize}
	\item The intensity reduction factor $t$.
	\item The restricted Lorentz transformation $\pmb{J}$: 
	\begin{itemize}
		\item Three parameters describing the rotation $\pmb{J}_{\mathrm{rot}}$.
		\item Three parameters describing the boost $\pmb{J}_{\mathrm{boost}}$.
	\end{itemize}
\end{itemize}
Incidentally, the matrix polar decomposition
guarantees that the Lorentz transformations can always be decomposed into a nonunique product of a single rotation and a single boost, as \cite{LuChipman1996}
\eq{
\pmb{J}=\pmb{J}_{\mathrm{rot}}\left(\Theta,\mathbf{n}_1\right)\pmb{J}_{\mathrm{boost}}\left(\eta,\mathbf{n}_2\right)= \pmb{J}_{\mathrm{boost}}\left(\eta^\prime,\mathbf{n}_2^\prime\right)\pmb{J}_{\mathrm{rot}}\left(\Theta,\mathbf{n}_1\right).
\label{eq:Jones polar}
} A deterministic polarization transformation, described by a single Jones matrix, can thus always be interpreted as resulting from an intensity reduction, a rotation, and a boost applied sequentially, in any order, to an incident beam of light's polarization degrees of freedom. 

We note that the composition of two rotations is another rotation, while the same does not hold true for two boosts. Instead, the product of two boosts becomes the composition of a boost and a rotation. This extra rotation is present in many physical situations \cite{Malykin2006,Tudor2018}, such as through the Thomas precession in special relativity \cite{Thomas1926} and the Wigner rotation in mathematical physics \cite{Wigner1939}.\footnote{The effect seems to have documented known before either Thomas or Wigner discovered their eponymous  effects \cite{Silberstein1914}.} This effect has indeed been investigated for light's polarization degrees of freedom \cite{Vigoureux1992,VigoureuxGrossel1993,MonzonSanchezSoto1999a,MonzonSanchezSoto1999b,MonzonSanchezSoto2001}, with the fascinating mathematical equivalence between indices of refraction for polarized light traveling through planar media and relative velocities of inertial frames in special relativity \cite{Vigoureux1992}.

\subsubsection{Deterministic Transformations in terms of Rotation and Diattenuation}
While the rotation transformations seen above are readily enacted by waveplates and liquid crystals, it is not immediately obvious to what laboratory equipment a Lorentz boost corresponds. 
Fortunately, the three boost parameters and the intensity reduction factor can be alternatively arranged into four ``diattenuation'' parameters, which is so named because 
diattenuations attenuate each of the field's polarization components by a different amount. 

The simplest example of a diattenuation is when the two circular components of polarization are each diminished by a different amount:
\eq{
\mathbf{J}_{\mathrm{diatten}}\left(q,r,\mathbf{e}_3\right)\begin{pmatrix}
    a\\b
\end{pmatrix}=\begin{pmatrix}
    a\sqrt{q}\\b\sqrt{r}
\end{pmatrix}.
\label{eq:Jones diattenuation}
} 
This transformation can arise from, for example, light reflecting off of a surface where there is a difference in reflectivity for electric fields that are parallel versus perpendicular to the plane of reflection \cite{Jackson1999}.
We can compare this transformation with Eqs. \eqref{eq:Jones scale factor} and \eqref{eq:Jones boost} to reveal the correspondences
\eq{
q&=t^2\eu^{\eta}\\
r&=t^2\eu^{-\eta} .
\label{eq:diattenuation parameters from boost etc}
} Since attenuations only serve to decrease the intensity of light in any mode, we can use the requirements $q,r\leq 1$ to provide physical constraints on the boost and overall transmission parameters from Eqs. \eqref{eq:Jones scale factor} and \eqref{eq:Jones boost}. For example, a polarizer that transmits only left-handed circularly polarized light but not its right-handed counterpart is described by the pair $\left(q,r\right)=\left(0,1\right)$; this then implies the limit of an infinite boost $\eta\to-\infty$ along the $\mathbf{e}_3$ axis, subject to the constraint $t=\exp\left(\eta/2\right)$.

We can also determine the Mueller matrices for arbitrary diattenuations using Eq. \eqref{eq:Mueller from pure Jones}. The above example of a boost along the $\mathbf{e}_3$ axis, with Jones matrix given by Eq. \eqref{eq:Jones diattenuation}, corresponds to the Mueller matrix
\eq{
\mathbf{M}_{\mathrm{diatten}}\left(q,r,\pmb{e}_3\right)=
\begin{pmatrix}
	\frac{q+r}{2} &0&0& \frac{q-r}{2}\\
	0&\sqrt{qr}&0&0\\
	0&0&\sqrt{qr}&0\\
	\frac{q-r}{2}&0&0&\frac{q+r}{2}
\end{pmatrix}.
\label{eq:Mueller diattenuation}
} Following a diattenuation, the total intensity $S_0$ decreases unless both $q$ and $r$ are equal to unity and all four Stokes parameters may decrease in magnitude. In the example of extreme attenuation corresponding to a polarizer, two components of $\mathbf{S}$ are completely nullified and the intensity of the retained component depends only on the original intensity of that component alone. The two polarizers to which our Eqs. \eqref{eq:Jones diattenuation} and \eqref{eq:Mueller diattenuation} may refer are
\eq{
\pmb{M}_{\mathrm{diatten}}\left(1,0,\mathbf{e}_3\right)=
\begin{pmatrix}
	\frac{1}{2} &0&0& \frac{1}{2}\\
	0&0&0&0\\
	0&0&0&0\\
	\frac{1}{2}&0&0&\frac{1}{2}
\end{pmatrix}\qquad \iff \qquad I_{\mathrm{R}}+I_{\mathrm{L}}\to I_{\mathrm{R}}}and \eq{ \pmb{M}_{\mathrm{diatten}}\left(0,1,\mathbf{e}_3\right)=
\begin{pmatrix}
	\frac{1}{2} &0&0& -\frac{1}{2}\\
	0&0&0&0\\
	0&0&0&0\\
	-\frac{1}{2}&0&0&\frac{1}{2}
\end{pmatrix}\qquad \iff \qquad I_{\mathrm{R}}+I_{\mathrm{L}}\to I_{\mathrm{L}}.
\label{eq:Mueller for polarizers}
}

The two other free parameters, in addition to $q$ and $r$, in a general diattentuation are the two angular coordinates of $\mathbf{n}$ dictating which two orthogonal modes are being attenuated. A general diattenuation is physically equivalent to first rotating the polarization such that the modes to be attenuated are $\mathrm{R}$ and $\mathrm{L}$, next applying the diattenuation given by Eqs. \eqref{eq:Jones diattenuation} and \eqref{eq:Mueller diattenuation}, and finally rotating the light back to its original polarization orientation. Even though a general rotation depends on three parameters, these rotations here depend only on the two parameters required to enact the rotation 
\eq{
\mathbf{R}\left(\mathbf{n}\to\mathbf{e}_3\right)\mathbf{n}=\mathbf{e}_3. 
} As such, more than one rotation is sufficient to describe this situation. Using any of these rotations, which are all orthogonal in the sense that $\mathbf{R}\left(\mathbf{n}\to\mathbf{e}_3\right)^\top=\mathbf{R}\left(\mathbf{n}\to\mathbf{e}_3\right)^{-1}=\mathbf{R}\left(\mathbf{e}_3\to\mathbf{n}\right)$, we can write the most general diattenuation transformation as
\eq{
\mathbf{M}_{\mathrm{diatten}}\left(q,r,\mathbf{n}\right)&=\mathbf{M}_{\mathrm{rot}}\left(\mathbf{n}\to\mathbf{e}_3\right)^{-1}
\mathbf{M}_{\mathrm{diatten}}\left(q,r,\mathbf{e}_3\right)
\mathbf{M}_{\mathrm{rot}}\left(\mathbf{n}\to\mathbf{e}_3\right)\\&=
\begin{pmatrix}
    1&\mathbf{0}^\top\\
    \mathbf{0}&\mathbf{R}\left(\mathbf{e}_3\to\mathbf{n}\right)
\end{pmatrix}
\begin{pmatrix}
	\frac{q+r}{2} &0&0& \frac{q-r}{2}\\
	0&\sqrt{qr}&0&0\\
	0&0&\sqrt{qr}&0\\
	\frac{q-r}{2}&0&0&\frac{q+r}{2}
\end{pmatrix}
\begin{pmatrix}
    1&\mathbf{0}^\top\\
    \mathbf{0}&\mathbf{R}\left(\mathbf{n}\to\mathbf{e}_3\right)
\end{pmatrix} .
\label{eq:Mueller diattenuation general}
}

Combining diattenuations with the matrix polar decomposition, any deterministic polarization transformation given by a single Jones matrix can be composed from a single rotation and a single diattenuation transformation, in either order \cite{LuChipman1996}. As the set of rotations forms a group under multiplication and inversion, we conclude that arbitrary deterministic polarization transformations can be obtained from waveplates and a single diattenuating element.

\subsubsection{Nondeterministic Transformations}
Deterministic transformations account for seven of the $16$ degrees of freedom of a general polarization transformation contained by the Mueller matrices of Eq. \eqref{eq:Mueller definition Stokes}.
The remaining nine parameters must arise from transformations requiring more than a single Jones matrix, such as through the convex combination in Eq. \eqref{eq:coherency multiple Jones}, which is what gives rise to the ``nondeterministic'' nomenclature.
Such transformations are also denoted as depolarizing because they almost always depolarize incident light that is perfectly polarized light (unless they simply do not effect any transformation on the latter). 

Nondeterministic polarization transformations are usually ascribed to the inability to experimentally discriminate between physical processes that each enact a different deterministic polarization transformation, as opposed to arising from fundamentally indeterminate laws of nature \cite{Gil2007}. For example, when a detector absorbs photons with a range of frequencies that are present in a quasimonochromatic beam of light and each frequency experiences a different polarization rotation after travelling through a birefringent crystal, the resultant polarization state must be taken to be the convex combination of the polarization states of the various frequency components, each with their individual polarization rotations \cite{LuLoeber1975,Loeber1982,Dlugnikov1984,Chakraborty1986}. Similarly, light whose polarization is rotated more quickly than can be resolved by a detector will have its degree of polarization reduced accordingly \cite{Billings1951}. Finally, light scattering off of optically active media has its polarization change depending on the direction of scattering \cite{Chakraborty1986}, so a detector receiving light from a nonzero range of solid angles leads to convex combinations of the deterministic processes ascribed to each scattering angle. These examples are well summarized by the assertion that that all depolarization processes can, in principle, be reversed but cannot be reversed in practice \cite{LuLoeber1975}.

The simplest example of a nondeterministic polarization transformation is that of an ideal depolarizer, whose Mueller matrix reads
\eq{
\mathbf{M}_{\mathrm{depol}}=\begin{pmatrix}
    1&0&0&0\\
    0&0&0&0\\
    0&0&0&0\\
    0&0&0&0
\end{pmatrix}.
} An ideal depolarizer maintains $S_0$, leaving the total intensity and total energy of the light unchanged, while enacting $p\to 0$, regardless of the input state. Equivalently, an ideal depolarizer transforms the perfectly polarized component of the beam represented in Eqs. \eqref{eq:Stokes decomposition} and \eqref{eq:coherency decomposition} into a completely unpolarized component. The polarization scrambling methods mentioned are made to simulate ideal depolarizers \cite{LuLoeber1975}, with work continuing to be done in this field of depolarizer design \cite{Zhangetal2007,Geetal2012,ShahamEisenberg2011,Marcetal2019,Krohetal2021,Marcoetal2021}.

Nondeterministic polarization transformations do not affect the total intensity $S_0$, only altering the polarization vector $\mathbf{S}$. The nine remaining parameters of general nondeterministic (``depolarizing'') Mueller matrices take the form of a $3\times 3$ symmetric real matrix $\mathbf{m}$ and a real vector $\mathbf{p}$ with magnitude less than or equal to unity \cite{LuChipman1996}:
\eq{
\mathbf{M}_{\mathrm{depol}}=\begin{pmatrix}
    1&\mathbf{0}^\top\\
    \mathbf{p} &\mathbf{m}
\end{pmatrix}.
} The assumption that only positive weights feature in Eqs. \eqref{eq:coherency multiple Jones with weights} and \eqref{eq:Mueller from Jones with weights} restricts the matrix $\mathbf{m}$ to being positive \cite{Gil2007}, which precludes transformations of the form of Eq. \eqref{eq:Mueller reflection}. Under some physical assumptions, such matrices can always be decomposed into the product of a single diattenuation, a single rotation, and a single depolarizing transformation with $\mathbf{p}=0$, through \cite{Gil2007}
\eq{
\mathbf{M}_{\mathrm{depol}}=\mathbf{M}_{\mathrm{rot}}\left(\Theta,\mathbf{n}\right)
\mathbf{M}_{\mathrm{diatten}}\left(q,r,\mathbf{n^\prime}\right)
\begin{pmatrix}
    1&\mathbf{0}^\top\\
    \mathbf{0} &\mathbf{m^\prime}
\end{pmatrix}.
}

How can we achieve an ideal depolarizer using this formalism? We note that, due to Eq. \eqref{eq:Mueller from Jones with weights}, nondeterministic Mueller matrices come from probabilistic mixtures of deterministic Mueller matrices
\eq{
\boldsymbol{\Psi}\to \sum_i \lambda_i \mathbf{J}_i\boldsymbol{\Psi} \mathbf{J}_i^\dagger \qquad \Rightarrow \qquad \mathbf{M}=\sum_i \lambda_i \mathbf{M}_i .
} 
If each of the deterministic transformations in the combination correspond to a rotation by angle $\Theta_i$ about axis $\mathbf{n}_i$, we achieve a depolarizing transformation with $\mathbf{p}=0$ and $\mathbf{m}=\sum_i \lambda_i \mathbf{R}\left(\Theta_i,\mathbf{n}_i\right)$. A sufficient number of rotations in a sufficient number of directions leads to $\mathbf{m}=\mathbf{0}\mathbf{0}^\top$. 
We further note that an \textit{arbitrary} symmetric matrix $\mathbf{m}$ can be obtained from a convex combination of sufficiently many rotation matrices with appropriate weights, provided that those weights are allowed to be negative \cite{Goldberg2020}. Otherwise, only positive matrices $\mathbf{m}$ can be generated.

We collect two few key facts before proceeding. First, any polarization transformation can be realized by the sequential application of a rotation, a diattenuation, and a depolarization, in any order \cite{LuChipman1996,Giletal2013}. Next, all such transformations can be realized via convex combinations of four or fewer deterministic transformations \cite{Cloude1986}. We thereby possess a complete description of all polarization transformations from a classical perspective.

\subsubsection{Physical Constraints on Polarization Changes}
\label{sec:classical constraints}
Many works have investigated physical constraints on the viability of various Jones and Mueller matrices \cite{FryKattawar1981, Simon1982, GilBernabeu1985, Cloude1990,vanderMee1993, Hovenier1994, Gil2000, Zakerietal2013, Cloude1986, LuChipman1996, Hovenieretal1986, Barakat1987, GivensKostinski1993, Nagirner1993, Simonetal2010, vanZyletal2011, Giletal2013}. We will not review all of them here, instead focusing on a few constraints that become relevant in the quantum theory of polarimetry.

Foremost, polarization changes are only considered to be physically viable if they are composed of probabilistic mixtures of pure Jones transformations, equivalent to Eqs. \eqref{eq:coherency multiple Jones with weights} and \eqref{eq:Mueller from Jones with weights} restricted to $\lambda_i>0$. This assertion can be traced to \textcite{Cloude1986}, which we will quote directly due to the challenge of obtaining this reference: ``What are the weighting coefficients [$\lambda_i$] and how are they determined?'' They proceed to answer this question ``by considering a new formulation of the scattering problem'' based on rearranging the elements of a rescaled Jones matrix $\pmb{J}$ into a $4\times 1$ vector $\mathcal{K}$. Then, a general Mueller matrix can be formed by taking linear combinations of the outer products $\mathbf{T}=\sum_i \lambda_i\mathcal{K}_i\mathcal{K}_i^\dagger$, eventually giving rise to Eq. \eqref{eq:Mueller from Jones with weights} with the eigenvalues of $\mathbf{T}$, $\lambda_i$, acting as the weights up to unitary transformations (``plane rotations in a 6 dimensional real target space''). This leads to the assertion that, because $\mathbf{T}$ ``is a complex correlation matrix [with] a much clearer physical interpretation than the Mueller matrix,'' it follows that ``the eigenvalues are positive real and each eigenvector corresponds . . . to a single scattering matrix.'' Physically, no reason is provided to prohibit a single Mueller matrix of the form of Eq. \eqref{eq:Mueller reflection} arising on its own in nature other than the, perhaps circular, perhaps physically reasonable, assertion that all Mueller matrices arise from probabilistic mixtures of deterministic transformations associated with single Jones matrices.

A flurry of attention revisited this problem motivated by quantum theory \cite{AhnertPayne2005,Aielloetal2007,Sudhaetal2008,Simonetal2010,GamelJames2011}, which we have conjectured to truly underlie the necessity of positive weights in Eqs. \eqref{eq:coherency multiple Jones with weights} and \eqref{eq:Mueller from Jones with weights} \cite{Goldberg2020}. It was first noted that the transformations responsible for transforming coherency matrices look like completely positive quantum channels, where the Jones matrices of Eq. \eqref{eq:coherency multiple Jones} can be thought of as Kraus operators \cite{AhnertPayne2005,Aielloetal2007,Sudhaetal2008,GamelJames2011}. This hints at the well-known connection with the quantum theory of polarization, viz., that single photons have density matrices described by their coherency matrices $\boldsymbol{\Psi}$; the transformations of classical polarization states are akin to transformations of single-photon polarization states. Then, all physically viable transformations that take single-photon polarization states to single-photon polarization states must be completely positive, immediately restricting the weights in Eqs. \eqref{eq:coherency multiple Jones with weights} and \eqref{eq:Mueller from Jones with weights} to be positive. We learn that, if all light were to be described by the behaviours of single photons (which precludes studies of attenuation), quantum theory would enforce the positivity assertion of classical polarization transformations. Then, inspired by a phenomenon known as ``nonquantum entanglement,'' \textcite{Simonetal2010} showed that the only way for an extended version of Eq. \eqref{eq:stokes squared inequality} to hold is through the positivity assertion of classical polarization transformations. In the extended version, the Stokes parameters are extended to two-point correlation functions, wherein the electric fields are taken to be at different points in the $x$-$y$ plane [cf. Eq. \eqref{eq:Stokes ensemble definition}]:
\eq{
    S_\mu\left(x,y;x^\prime,y^\prime\right)=\frac{1}{2}\expct{\mathbf{E}\left(x^\prime,y^\prime,z;t\right)^\dagger\sigma_\mu \mathbf{E}\left(x,y,z;t\right)}.
    \label{eq:two point Stokes parameters}
} In order to act linearly on these extended Stokes parameters, Mueller matrices must be made from positive-weight combinations of deterministic elements. Of course, this is asking more of Mueller matrices than the standard definition in Eq. \eqref{eq:Mueller definition Stokes}, so it remains to be proven whether the necessity of positive weights in Eqs. \eqref{eq:coherency multiple Jones with weights} and \eqref{eq:Mueller from Jones with weights} follows from any deeper physical condition.

The most natural constraint on Mueller and Jones matrices is that they take physically viable polarization states to physically viable polarization states. This means that all linear polarization transformations must ensure their output states satisfy the constraint of Eq. \eqref{eq:stokes squared inequality} regardless of the input state. All pure Jones matrices automatically satisfy this constraint, as do convex combinations thereof, so we learn that the 16 parameters of a polarization transformation are not independently free, instead satisfying \eq{\mathrm{Tr}\left(\mathbf{M}\mathbf{M}^\top\right)\leq 4 M_{0,0}^2.}

Finally, we mention the transmittance and reverse transmittance conditions for Mueller matrices \cite{Gil2000}. 
Under the assumption that a single deterministic transformation does not increase the intensity of an incident beam, 
it can be verified that any Mueller matrix arising from a single deterministic transformation satisfies
\eq{
    M_{0,0}+\sqrt{M_{0,1}^2+M_{0,2}^2+M_{0,3}^2}=M_{0,0}+\sqrt{M_{1,0}^2+M_{2,0}^2+M_{3,0}^2}\leq 1\,\, \iff\,\, \mathrm{Tr}\left(\mathbf{M}\mathbf{M}^\top\right)= 4 M_{0,0}^2.
} Then, any Mueller matrix arising from a convex combination of deterministic elements must satisfy
\eq{
    M_{0,0}+\sqrt{M_{0,1}^2+M_{0,2}^2+M_{0,3}^2}&\leq 1\qquad\mathrm{and}\qquad
    M_{0,0}+\sqrt{M_{1,0}^2+M_{2,0}^2+M_{3,0}^2}&\leq 1\\ &\qquad\iff\qquad \lambda_i>0\mathrm{\,in\, Eqs.\, \eqref{eq:coherency multiple Jones with weights}\, and\, \eqref{eq:Mueller from Jones with weights}}.
} These conditions preclude the possibility of transformations such as lossless polarizers, which would be able to convert all input light of a given polarization component into light with the opposite component regardless of input beam, such as
\eq{
    \mathbf{M}_{\mathrm{lossless\,polarizer}}=\begin{pmatrix}
        1&0&0&0\\
        0&0&0&0\\
        0&0&0&0\\
        1&0&0&0
    \end{pmatrix}.
    \label{eq:lossless polarizer Mueller}
}

It turns out that we can concoct polarization transformations that satisfy all of the requisite conditions, even that of positive weights, while breaking the transmittance conditions. For example, we can consider the pair of pure Jones matrices
\eq{
    \mathbf{J}_1=\begin{pmatrix}
        1&0\\0&0
    \end{pmatrix}\qquad\mathrm{and}\qquad \mathbf{J}_2=\begin{pmatrix}
        0&1\\0&0
    \end{pmatrix},
    \label{eq:lossless polarizer Jones}
} which directly lead to the Mueller matrix of Eq. \eqref{eq:lossless polarizer Mueller} when added with $\lambda_1=\lambda_2=1$ as in Eq. \eqref{eq:coherency multiple Jones} (similarly, if we employ $\mathbf{J}_2^\top$ instead of $\mathbf{J}_2$, we find $\mathbf{M}_{\mathrm{lossless\,polarizer}}^\top$, which disobeys the forward instead of the reverse transmittance condition). On the contrary, if we consider adding them with $\lambda_1=\lambda_2=1/2$ as in Eq. \eqref{eq:coherency multiple Jones with weights}, we find the resulting Mueller matrix to be $\mathbf{M}=\mathbf{M}_{\mathrm{lossless\,polarizer}}/2$, which indeed satisfies the reverse transmittance condition. How should we rectify this situation? Requiring the coefficients $\lambda_i$ to sum to unity would suffice, because the transmittance conditions assume we cannot rescale these Jones matrices by a factor greater than unity. In contradistinction, the unital nature of quantum channels only requires Kraus operators to satisfy $\mathbf{J}_1^\dagger \mathbf{J}_1+\mathbf{J}_2^\dagger \mathbf{J}_2=\mathds{1}$, which is indeed satisfied here with $\lambda_1=\lambda_2=1$, so there seems to exist a quantum channel that would act like a lossless polarizer on a single photon. Further inspection reveals that $\mathbf{J}_1$ ensures that all right-handed circularly polarized light retains its polarization and $\mathbf{J}_2$ converts all left-handed circularly polarized light to its right-handed counterpart. This can only be achieved in a ``linear'' manner by a device that measures the total incident intensity, then prepares a right-handed circularly polarized state with that same intensity to be output: a highly nonlinear device poising as a linear one. Because this must hold regardless of the input Stokes parameters, it must be capable of measuring and generating output states of light with arbitrarily large intensities. These considerations prohibit such a device from being physically viable and reinforce the transmittance conditions; we will return to this consideration when discussing quantum polarization transformations.


\section{Quantum Polarization}
\label{sec:quantum polarization}
Maxwell's equations equally apply to quantized light fields. In the quantum theory, the field amplitudes $a$ and $b$ from Eq. \eqref{eq:plane wave} get promoted to bosonic operators $\ha$ and $\hb$ that annihilate right- and left-handed circularly polarized photons:
\eq{
    \hat{\mathbf{E}}=\mathcal{E}_0\left(\ha\mathbf{e}_a+\hb\mathbf{e}_b\right)\eu^{\iu\left(kz-\omega t\right)}.
} These operators satisfy the standard bosonic commutation relations
\eq{
    \left[\ha,\had\right]=\left[\hb,\hbd\right]=1\qquad \mathrm{and}\qquad\left[\ha,\hbd\right]=\left[\hb,\had\right]=0
} and
can be used to create states with definite photon numbers in a single spatial mode from the two-mode vacuum $\vac$ via
\eq{
    \ket{m,n}\equiv \ket{m}_{\mathrm{R}}\otimes\ket{n}_{\mathrm{L}}=\frac{\had^m\hbd^n}{\sqrt{m!n!}}\vac.
} The intensity of the field is given by the average number of excitations $\langle \had \had+\hbd\hb\rangle$ and similarly for the intensities within each polarization component.

Following the above quantization rule, we can define the Stokes operators as
\eq{
    \hat{S}_0&=\frac{\hat{n}_{\mathrm{H}}+\hat{n}_{\mathrm{V}}}{2}=\frac{\hat{n}_{\mathrm{D}}+\hat{n}_{\mathrm{A}}}{2}=\frac{\hat{n}_{\mathrm{R}}+\hat{n}_{\mathrm{L}}}{2}=\frac{\had\ha+\hbd\hb}{2},\\
    \hat{S}_1&=\frac{\hat{n}_{\mathrm{H}}-\hat{n}_{\mathrm{V}}}{2}=\frac{\had\hb+\hbd\ha}{2},\\
    \hat{S}_2&=\frac{\hat{n}_{\mathrm{D}}-\hat{n}_{\mathrm{A}}}{2}=-\iu\frac{\had\hb-\hbd\ha}{2},\\
    \hat{S}_3&=\frac{\hat{n}_{\mathrm{R}}-\hat{n}_{\mathrm{L}}}{2}=\frac{\had\ha-\hbd\hb}{2},
    \label{eq:Stokes operators}
}
or succinctly through [cf. Eq. \eqref{eq:Stokes ensemble definition}]
\eq{
    \hat{S}_\mu=\frac{1}{2}\begin{pmatrix}
        \had &\hbd
    \end{pmatrix}\sigma_\mu\begin{pmatrix}
        \ha \\\hb
    \end{pmatrix},
}
with the quantum-to-classical correspondence
\eq{
    S_\mu=\expct{\hat{S}_\mu}.
} The factor of $2$ that we have been carrying through these definitions allows us to realize the Stokes operators as obeying the commutation relations of angular momentum,
\eq{
    \left[\hat{S}_i,\hat{S}_j\right]=\sum_{k=1}^3\epsilon_{i j k}\hat{S}_k\qquad \mathrm{and}\qquad \left[\hat{S}_0,\hat{S}_i\right]=0,
} and we see that the Schwinger mapping \cite{Chaturvedietal2006,SakuraiNapolitano2011} governs the translation between two-mode states and angular momentum eigenstates.

Classically, all of the Stokes parameters can be measured with arbitrary precision. However, since they arise as expectation values of noncommuting operators, the same cannot be said at a fundamental level; instead, there is a lower limit to the joint precision with which a pair of Stokes parameters can be measured. Methods of optimizing tradeoffs such as
\eq{
    4\Var \hat{S}_1\Var \hat{S}_2\geq \left|\expct{\hat{S}_3}\right|^2
}
and
\eq{
    \Var \hat{S}_1+\Var \hat{S}_2+\Var \hat{S}_3\geq \expct{\hat{S}_0},
    \label{eq:Stokes variance sum inequality}
} where we denote operator variances by $\Var X=\expct{X^2}-\expct{X}^2$,
have led to the fruitful discovery and implementation of ``polarization squeezing,'' which may ultimately have applications in quantum-enhanced polarimetry and other communication tasks \cite{Winelandetal1992,Chirkinetal1993,KitagawaUeda1993,HilleryMlodinow1993,AgarwalPuri1994,Winelandetal1994,Karassiov1994,Alodzhantsetal1995,BrifMann1996,Klyshko1997,Alodjantsetal1998,Haldetal1999,SorensenMolmer2001,Korolkovaetal2002,Bowenetal2022,AbdelAtyetal2002,Schnabeletal2003,Heersinketal2003,Josseetal2003,AndersenBuchhave2003,Shindoetal2003,WangSanders2003,Golubevetal2004,Heersinketal2005,KorolkovaLoudon2005,Popescu2005,LuisKorolkova2006,ShersonMolmer2006,Marquadtetal2007,Chaudhuryetal2007,Lassenetal2007,Milanovicetal2007,RivasLuis2008,Corneyetal2008,Dongetal2008,Iskhakovetal2009,Shalmetal2009,GuhneToth2009,Hsuetal2009,Mahleretal2010,Milanovicetal2010,Heetal2011,PrakashShukla2011,Maetal2011,Barreiroetal2011,Gross2012,Puentesetal2013,Civitareseetal2013,KlimovMunoz2013,BeduiniMitchell2013,MitchellBeduini2014,Chirkin2015,Puentes2015,Mulleretal2016,Vitaglianoetal2017,Wenetal2017,Hanetal2018,Vitaglianoetal2018,Birrittellaetal2021,Baietal2021}. Polarization squeezing is significantly reviewed in Refs \cite{Maetal2011}, with briefer reviews in, e.g., Refs. \cite{TothApellaniz2014,Chirkin2015,Goldbergetal2021polarization}.

From the theory of angular momentum, we immediately find that
\eq{
    \hat{\mathbf{S}}^2\equiv \hat{S}_1^2+\hat{S}_2^2+\hat{S}_3^2=\hat{S}_0\left(\hat{S}_0+1\right)\geq \hat{S}_0^2.
} This does not imply that quantum states may disobey criterion of Eq. \eqref{eq:stokes squared inequality}, which remains true in the guise of
\eq{
    \expct{\hat{S}_1}^2+\expct{\hat{S}_2}^2+\expct{\hat{S}_3}^2\leq \expct{\hat{S}_0}^2;
} these together enforce the variance inequality in Eq. \eqref{eq:Stokes variance sum inequality}. Interestingly, the ultimate quantum limit with which the Stokes parameters \textit{for classical light} may be simultaneously estimated obeys stricter conditions than Eq. \eqref{eq:Stokes variance sum inequality}:
\eq{
    \Var \hat{S}_1+\Var \hat{S}_2+\Var \hat{S}_3\geq \frac{5}{2}\expct{\hat{S}_0}
    \label{eq:uncertainty limit classical state 1}
} when the value of $S_0$ is unknown \cite{Kikuchi2020,Mecozzietal21,Jarzyna2021arxiv} and
\eq{
    \Var \hat{S}_1+\Var \hat{S}_2+\Var \hat{S}_3\geq 2\expct{\hat{S}_0}
    \label{eq:uncertainty limit classical state 2}
} when the value of $S_0$ is known \textit{a priori} \cite{Jarzyna2021arxiv}.

The four Stokes parameters can be simultaneously measured by mixing the input light with six other vacuum modes input to an interferometer and computing sum or difference currents among four particular pairs of output modes \cite{AlodjantsArakelian1999,Alodjantsetal1999}. Each sum or difference photocurrent yields one half of one of the Stokes parameters, even in the quantum regime, but the variances of these photocurrents are not proportional to the variances of the Stokes parameters; instead, the former are all offset from the latter by an amount proportional to the total intensity of the field and are thus never nonzero.
Similar schemes can be created using interferometers with two \cite{LuisPerina1996,Richter2000} or four \cite{LuisPerina1996,RivasLuis2008covariance} input ports in their vacuum states.
Mathematical postprocessing of such a measurement can be used to verify the presence of polarization squeezing and the potential saturation of inequalities such as Eq. \eqref{eq:Stokes variance sum inequality}.

It is instructive to investigate the properties of single-photon polarization states, spanned by $\ket{1,0}$ and $\ket{0,1}$, as these directly encompass classical polarization phenomena, which are governed by the $2\times 2$-density-matrix-like object $\boldsymbol{\Psi}$. The most general density matrix for such a state is given in this basis by
\eq{
    \hat{\rho}_{\mathrm{single\,photon}}=\begin{pmatrix}
        \rho_{0,0}&\rho_{1,0}^*\\
        \rho_{1,0}&1-\rho_{0,0}
    \end{pmatrix},
} subject to the constraints of positivity. The Stokes parameters are readily calculable using Pauli matrices, yielding
\eq{
    \mathcal{S}=\begin{pmatrix}
        \frac{1}{2}\\
        \RE \rho_{1,0}\\
        \IM \rho_{1,0}\\
        \rho_{0,0}-\frac{1}{2}
    \end{pmatrix}.
} The density matrix can be decomposed into density matrices corresponding to pure and maximally mixed states, via
\eq{
    \hat{\rho}_{\mathrm{single\,photon}}=\sum_{\mu=0}^3 S_\mu\sigma_\mu=p\frac{\mathds{1}+\frac{\mathbf{S}}{\left|\mathbf{S}\right|}\cdot\boldsymbol{\sigma}}{2}+\left(1-p\right)\frac{\mathds{1}}{2},
    \label{eq:rho single photon from Stokes}
} where the degree of polarization is, as usual, $p=\left|\mathbf{S}\right|/S_0$. Pure single photons have degree of polarization $p=1$ and maximally mixed single photons have $p=0$, which can alternatively be expressed via the purity parameter $\Tr \left(\hat{\rho}^2\right)$ through
\eq{
    p=\frac{\left|\mathbf{S}\right|}{S_0}=\sqrt{\frac{\Tr\left(\boldsymbol{\Psi}^2\right)}{2S_0^2}-1}=\sqrt{2\Tr\left(\hat{\rho}^2\right)-1}.
    \label{eq:purity polarization single photons}
} The density matrix for single photons thus completely reproduces the classical coherency matrix for describing polarization states. We next explore properties of quantum states with more than a single photon.

\subsection{Characterizing Polarization}
The Stokes operators presented in Eq. \eqref{eq:Stokes operators} conserve photon number, as each creation operator is paired with an annihilation operator. This is one way to see why they commute with the total-photon-number operator $\hat{N}=2\hat{S}_0$ and ensures that the polarization properties of a beam of light can be broken into the polarization properties of each photon-number subspace, where the latter are sometimes called Fock layers \cite{Donatietal2014,Mulleretal2016}. We next explore the polarization properties of pure states with a fixed number of photons $N$, equivalent to states with spin $N/2$, whose most general form is given by
\eq{
    \ket{\psi^{(N)}}=\sum_{m=0}^N\psi_m\ket{m,N-m},\qquad \sum_{m=0}^N \left|\psi_m\right|^2=1.
    \label{eq:pure state Nth layer in terms of psim}
}

Geometrically, it is easy to visualize pure single-photon states on the surface of the Poincar\'e (or Bloch) sphere, as they can be parametrized by two angular coordinates $\Omega=\left(\Theta,\Phi\right)$:
\eq{
    \ket{\Omega^{(1)}}=\cos\frac{\Theta}{2}\ket{1,0}+\eu^{\iu\Phi}\sin\frac{\Theta}{2}\ket{0,1}.
} We can alternatively write these states as resulting from the action of a single creation operator, parametrized by $\Omega$, acting on the two-mode vacuum, through
\eq{
    \ket{\Omega^{(1)}}=\ha_\Omega^\dagger\vac,\qquad \ha_\Omega^\dagger=\cos\frac{\Theta}{2}\had+\eu^{\iu\Phi}\sin\frac{\Theta}{2}\hbd.
} However, we cannot visualize an $N$-photon state as a series of $N$ Poincar\'e spheres, as the photons are, in general, mutually correlated. This challenge is addressed by the Majorana representation \cite{Majorana1932,BengtssonZyczkowski2017}.
 
\begin{figure*}
    \centering
    \includegraphics[width=\textwidth]{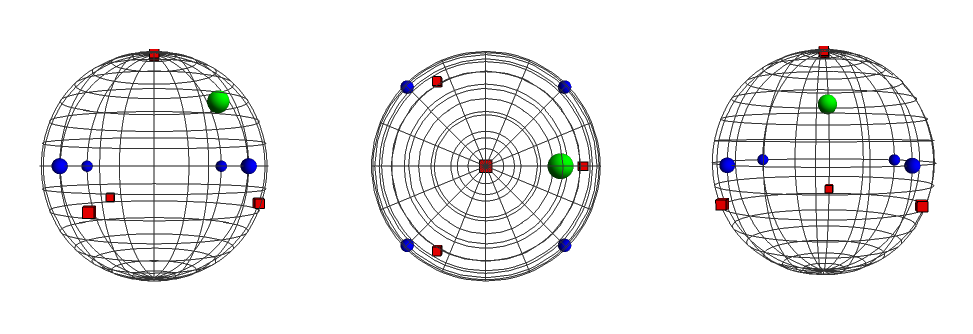}
    \caption{Three views of three different Majorana constellations for four-photon states. The large green ball corresponds to the fourfold degenerate constellation of an SU(2)-coherent state with $\Omega_1=\Omega_2=\Omega_3=\Omega_4$, the smaller blue balls correspond to a NOON state with Majorana stars equally spread about the equator, and the red cubes correspond to a state whose Majorana constellation is a regular tetrahedron.}
    \label{fig:Majorana examples}
\end{figure*}

The Majorana representation begins by realizing the one-to-one correspondence between the amplitudes $\left\{\psi_m\right\}$ and the set of $N$ angular coordinates $\left\{\Omega_k\right\}$, through
\eq{
    \ket{\psi^{(N)}}=\frac{1}{\sqrt{\mathcal{N}}}\prod_{k=1}^N \ha_{\Omega_k}^\dagger\vac,
} where the normalization constant $\mathcal{N}$ depends on the coordinates and does not affect the geometry of the state. This correspondence allows us to represent any $N$-photon pure state by a constellation of $N$ points on the surface of a sphere, where each point is sometimes referred to as a star in the Majorana constellation.\footnote{The constellation is often taken to be the set of points antipodal to these $N$ angular coordinates $\Omega_k$ so as to correspond directly to the zeroes of the Husimi $Q$-function.} Notably, the entire constellation rotates rigidly under a polarization rotation, lending geometric intuition to multiphoton polarization states. The usefulness of the Majorana representation has been realized in topics from metrology
\cite{ChryssomalakosHC2017,Bouchardetal2017,GoldbergJames2018Euler,Martinetal2020,Goldberg2020,Goldbergetal2021rotationspublished,Chryssomalakosetal2021} to Bose-Einstein condensates \cite{Lianetal2012,Cuietal2013} to non-Hermitian physics \cite{Bartlettetal2021} and beyond
\cite{Hannay1998JPA,Hannay1998JMO,Bjorketal2015,Bjorketal2015PRA,Giraudetal2010,KolenderskiDemkowiczDobrzanski2008,MakelaMessina2010PS,Martinetal2010,Lamacraft2010,Bruno2012,UshaDevietal2012,Yangetal2015,LiuFu2016,Chabaudetal2020,Dograetal2020}. We exemplify some Majorana constellations in Fig. \ref{fig:Majorana examples} and note that states whose Majorana constellations are randomly distributed have intriguing properties that are only now being elucidated \cite{Goldbergetal2021randomarxiv}.

The Majorana constellation, as it stands, pertains to pure $N$-photon states. What can we retain when considering more general quantum states? For mixed states with a fixed number of photons, representations have been derived that retain some of the geometrical properties of the standard constellation by either decomposing the density matrix into its eigenbasis \cite{Migdal2011} or into the spherical tensor basis \cite{SerranoEnsastigaBraun2020} and finding a constellation for each element in said basis. For pure states with indeterminate numbers of photons, one can consider a Majorana representation within each photon-number subspace, where one must also keep track of the relative weights and relative phases between each subspace (ignoring the relative phases, one can consider a set of Majorana constellations for a convex combination of pure states that each have a different number of photons) \cite{Bjorketal2015}. Regrettably, a unified geometrical picture for arbitrary quantum states \`a la Majorana is still lacking.

\subsubsection{Polarized States}
We are now in a position to ascertain which quantum mechanical states underlie their classical counterparts with degree of polarization $p=1$ \cite{GoldbergJames2017}, expanding upon earlier work by \textcite{MehtaSharma1974,PrakashSingh2000,SinghPrakash2013,Luis2016}. These quantum states have significant complementarity properties \cite{Norrmanetal2020}.
We note in passing that other degrees of polarization have been proposed in light of the quantum nature of polarization
\cite{Alodjantsetal1998,AlodjantsArakelian1999,Klimovetal2010,Luis2002,Luis2003,Luis2007PRAtypeII,Luis2016,SanchezSotoetal2006,Luis2007OptComm,Bjorketal2010,Ghiuetal2010,Ghiuetal2018,delaHozetal2013,Kotheetal2013,Bjorketal2012,SanchezSotoetal2013}, each with their own merits and motivations, but continue our investigation along the lines of the canonical degree of polarization because the proposed new degrees are mutually inconsistent in their orderings of partially polarized states \cite{Ghiuetal2018}.

The easiest example of a perfectly polarized quantum state is that of a pure single photon. This is readily generalized to pure states with exactly $N$ photons, where the resulting state is completely polarized if and only if all $N$ of the constituent photons have the same direction of polarization. In terms of the Majorana constellation, this requires all $N$ of the angular coordinates to degenerate to a single point on the surface of the sphere as in Fig. \ref{fig:Majorana examples}, with the angular coordinates of that point dictating the state's polarization direction. These states are the well-known 
SU(2)-coherent, or spin-coherent, states \cite{Arecchietal1972}
\eq{\ket{\Omega^{(N)}}=\frac{\ha_{\Omega}^\dagger\vphantom{a}^N}{\sqrt{N!}}\vac=\sum_{m=0}^N \psi_m^{\Omega;N}\ket{m,N-m},
\label{eq:SU(2) coherent states}
}where
\eq{\psi_m^{\Omega;N}=\sqrt{\binom{N}{m}}\cos^m\frac{\Theta}{2}\sin^{N-m}\frac{\Theta}{2}\eu^{\iu\Phi(N-m)}. \label{eq:coherent state amplitudes}
} SU(2)-coherent states
are eigenstates of an angular momentum operator projected in the direction $\mathbf{n}_{\Omega}=\left(\sin\Theta\cos\Phi,\sin\Theta\sin\Phi,\cos\Theta\right)^\top$ with eigenvalue $S_0=N/2$: 
\eq{\left(\hat{\mathbf{S}}\cdot\mathbf{n}_{\Omega}\right)\ket{\Omega^{(N)}}=\frac{N}{2}\ket{\Omega^{(N)}}
\label{eq:Stokes from SU(2) coherent}
} and have many other useful properties \cite{Perelomov1986,Gazeau2009}.
The Stokes parameters for SU(2)-coherent states are exactly those of perfectly polarized light with intensity equal to that of $N$ photons
\eq{\mathcal{S}=\frac{N}{2}\begin{pmatrix}
    1\\\mathbf{n}_{\Omega}
\end{pmatrix}}
and SU(2)-coherent states can be considered as spin states with maximal spin projection. Notably, these perfectly polarized states behave sensibly when undergoing polarization rotations, with the direction of polarization rotating as expected for classical beams of light.

The only states, pure or mixed, with exactly $N$ photons that are perfectly polarized are the SU(2)-coherent states of Eq. \eqref{eq:SU(2) coherent states}. The rest of the quantum states with $p=1$ must therefore have indeterminate photon number (i.e., not be eigenstates of $\hat{S}_0$).

One simple extension of SU(2)-coherent states is to convex combinations of SU(2)-coherent states, each with the same angular coordinates.
One can prove that such states, given by 
\eq{
\hat{\rho}=\sum_{N=0}^\infty \rho_N \ket{\Omega^{(N)}}\bra{\Omega^{(N)}},\quad \sum_{N=0}^\infty \rho_N=1, \quad \rho_N\geq 0 ,
\label{eq:convex combos of spin coherent states}
} are all perfectly polarized in direction $\Omega$, with intensity equal to the average photon number (in the appropriate units)
$\hat{S}_0\propto \sum_{N=0}^\infty\rho_N N$ \cite{MehtaSharma1974}. Even though there is a probabilistic mixture present, all of the states in the mixture have the same direction of polarization, so they conspire to yield a state that is completely polarized overall.
Equation \eqref{eq:convex combos of spin coherent states} is markedly different from the case of single photons: for single photons, purity and degree of polarization are the same quantity, as seen in Eq. \eqref{eq:purity polarization single photons}; when more than one photon number is involved, mixed states can still have degree of polarization $p=1$.

Returning momentarily from mixed states back to pure states,
superpositions of SU(2)-coherent states in different photon-number subspaces whose directions of polarization are all collinear are also perfectly polarized. In fact, these are the only possible pure states with degree of polarization $p=1$ \cite{GoldbergJames2017}:
\eq{
\ket{\Psi}=\sum_{N=0}^\infty\eu^{\iu\varphi_N}\sqrt{\rho_N}\ket{\Omega^{(N)}},\quad \sum_{N=0}^\infty \rho_N=1, \quad \rho_N\geq 0,\quad\varphi_N\in\mathds{R}.
\label{eq:pure perfectly polarized}
}
This simply means that classical polarization properties may be underlain by quantum superpositions about which the former are ignorant, especially because even the Stokes operators themselves do not distinguish between the pure superpositions of Eq. \eqref{eq:pure perfectly polarized} and the mixed states of Eq. \eqref{eq:convex combos of spin coherent states} in any of their correlation functions. In that sense, another form of correlation gadget is required to be sensitive to these relative phases between Fock layers, as outlined in the conclusions of \cite{Goldbergetal2021multipolesarxiv}, such as by using weak-field homodyne detection \cite{Donatietal2014}.

The pure states we are now discussing encompass canonical coherent states, which are generally agreed to be the most classical states according to quantum optics\cite{MandelWolf1995}:
\eq{
\ket{\alpha}\propto \exp(\alpha\had)\vac.
\label{eq:canonical coherent state}
} These states obey the restrictions of Eqs. \eqref{eq:uncertainty limit classical state 1} and \eqref{eq:uncertainty limit classical state 2} in terms of their simultaneously measurable properties.
From the perspective of polarization, these states take the form \cite{AtkinsDobson1971} \eq{\ket{\alpha_{\Omega}}\propto \exp(\alpha\ha_{\Omega}^\dagger)\vac =\sum_{N=0}^\infty \frac{\alpha^N}{N!}\ha_{\Omega}^\dagger\vphantom{a}^N\vac=\sum_{N=0}^\infty\frac{\alpha^N}{\sqrt{N!}}\ket{\Omega^{(N)}}.
\label{eq:canonical coherent states any pol}
} 
These are the states sometimes thought to underlie classical polarization phenomena, as they can be described solely using the Stokes vector [cf. Eqs. \eqref{eq:stokes pol} and \eqref{eq:Stokes from SU(2) coherent}]
\eq{
    \mathcal{S}=\frac{\left|\alpha\right|^2}{2}\begin{pmatrix}
        1\\
        \mathbf{n}_\Omega
    \end{pmatrix}
} with no other free parameters (i.e., no additional ``quantum'' degrees of freedom present beyond the classical description), but we will see later that even this thinking has its pitfalls when we begin to consider partially polarized states.

We are now in the position to write the most general perfectly polarized quantum state:
\eq{
\hat{\rho}=\sum_{M,N=0}^\infty \sigma_{M,N} \ket{\Omega^{(M)}}\bra{\Omega^{(N)}},\quad \Tr\pmb{\sigma}=1,
} for any positive-semidefinite matrix $\pmb{\sigma}$. These states can be thought of as probabilistic mixtures of pure states of the form of Eq. \eqref{eq:pure perfectly polarized}, which can also be realized using a single element from a pair of orthogonal creation operators via:
\eq{\hat{\rho}=\sum_{i=0}^\infty F_i\left(\ha_{\Omega}^\dagger\right)\vac\bra{\mathrm{vac}}F_i^*\left(\ha_{\Omega}\right),}
where the functions $F_i(z)=\sum_N f_N^{(i)}z^N$ need only be normalized by a common factor.
Moreover, these states arise exclusively from polarization rotations of states that have all of their excitations in a single mode, using the polarization rotation operators $\hat{R}$ that we will discuss in Section \ref{sec:quantum rotations}:
\eq{\hat{\rho}=\hat{R}\left(\hat{\sigma}_{\mathrm{R}}\otimes \ket{0}_{\mathrm{L}}\bra{0}\right)\hat{R}^\dagger.}
While a formal proof of these facts can be found in \cite{GoldbergJames2017}, we presented a simpler proof in \cite{Goldberg2021thesis} that only relies on polarization rotations and that
\eq{
    \expct{\hbd\hb}=0\qquad\iff\qquad\hat{\rho}=\hat{\sigma}_{\mathrm{R}}\otimes \ket{0}_{\mathrm{L}}\bra{0}.
}

In summary, perfectly polarized states have all of their photons conspire to seem completely classical. 
We stress, still, that the purity of such states can be quite low, in contradistinction to our classical intuition. We are now positioned to discuss the other term in the decompositions of Eq. \eqref{eq:Stokes decomposition} and \eqref{eq:coherency decomposition}, corresponding to unpolarized states, to partner with the perfectly polarized states and complete our quantum description of classical polarization.

\subsubsection{Unpolarized States}
\label{sec:unpolarized states}
Classically, unpolarized states are those that are unchanged by polarization rotations and such states have $p=0$.
Quantum mechanically, there is a marked difference between states unchanged by polarization rotations and states with $p=0$. This strongly underscores the differences between classical and quantum intuition in the realm of polarization.

Quantum states with $p=0$ must have $\expct{\hat{\mathbf{S}}}=\mathbf{0}$. This imposes three constraints onto a generic state that has many more than three degrees of freedom, so it is not surprising than many different states may underly classically unpolarized light. 

We begin with unpolarized single-photon states. These are given by the density matrices from Eq.  \eqref{eq:rho single photon from Stokes} with $\mathbf{S}=\mathbf{0}$ and correspond to maximally mixed states, according with the purity condition of Eq.
\eqref{eq:purity polarization single photons}. No free parameters remain, so all unpolarized single-photon states are the same, without any extra quantum mechanical degrees of freedom.

Unpolarized pure states of $N$ photons must satisfy the constraints
\eq{
    \sum_{m=0}^N\left|\psi_m\right|^2=1,\quad 2\sum_{m=0}^N m \left|\psi_m\right|^2=N,\quad \sum_{m=1}^N\psi_m^*\psi_{m-1}\sqrt{m\left(N-m+1\right)}=0.
} These four constraints, including normalization, can be compared to the free parameters of an $N$ photon state: the latter has $N+1$ complex degrees of freedom subject to normalization and the irrelevance of a global phase. Such unpolarized states thus retain $2N-3$ degrees of freedom, in stark contrast to the classical picture that cannot distinguish between any of these degrees of freedom. In addition, this directly shows that all unpolarized pure states of light must have more than one photon, signifying that one must look beyond $2\times 2$ matrices for investigating all polarization phenomena.

It is easy to geometrically concoct quantum states that are unpolarized with $p=0$ by taking advantage of symmetry properties through the Majorana representation. For example, the so-called NOON states are given by superpositions of SU(2)-coherent states pointed in opposite directions [$\Omega_\perp=\left(\Pi-\Theta,\Phi+\Theta\right)$]
\eq{
    \ket{\psi_{\mathrm{NOON}}}=\frac{\ket{\Omega^{(N)}}+\ket{\Omega_\perp^{(N)}}}{\sqrt{2}};
    \label{eq:NOON}
} such states have Majorana constellations equally spread about a single great circle, as depicted along the equator in Fig. \ref{fig:Majorana examples}. Similarly, one can consider states whose Majorana constellations have three-dimensional symmetry, including those corresponding to platonic solids like the tetrahedron (again, see Fig. \ref{fig:Majorana examples})
\eq{
    \ket{\psi_{\mathrm{tet}}}=\frac{\ket{4,0}+\sqrt{2}\ket{1,3}}{\sqrt{3}}\propto\had\left(\had+\sqrt{2}\hbd\right)\left(\had+\eu^{\iu2\pi/3}\sqrt{2}\hbd\right)\left(\had+\eu^{\iu4\pi/3}\sqrt{2}\hbd\right)\vac.
} Since Majorana constellations rotate rigidly under polarization rotations, any such rotation will preserve the unpolarized nature of a state.

We can also use the geometrical picture without much elegance by considering unpolarized states to be, for example, superpositions of SU(2)-coherent states pointed in opposite directions in different photon-number subspaces. A straightforward example is a state such as
\eq{
    \ket{\psi_{\mathrm{NOON-inspired}}}=\frac{\sqrt{N}\ket{\Omega^{(M)}}+\sqrt{M}\ket{\Omega_\perp^{(N)}}}{\sqrt{M+N}}.
} This should make it clear that there are an infinite number of possible quantum states with classical degree of polarization $p=0$, even though the classical picture assumes them all to be identical. Moreover, we have only shown an infinite number of possible pure unpolarized states; any convex combination of such states will also be unpolarized, so there are a plethora of unpolarized mixed states according to the classical degree.

A final note about classically unpolarized states is due to \textcite{Klyshko1992}. Orthogonal states such as $\ket{1,1}$ and $\left(\ket{2,0}+\ket{0,2}\right)/\sqrt{2}$ seem to be unpolarized, but they have the same Majorana constellation up to a rigid rotation (two antipodal points), so they can be interconverted via polarization rotations. The same is true of any pure $N$-photon state with odd number $N$ \cite{Sehatetal2005}. This is terrible from the perspective of classical polarization: polarization rotations should not change the measurable properties of a state, yet, somehow, a polarization rotation is here converting a state into an orthogonal one, which can readily be distinguished from the former. The pitfalls of classical polarization intuition continue to be elucidated.

An alternative to states simply satisfying $p=0$ is the set of states that are completely unchanged by polarization rotations. These states were shown by \textcite{Agarwal1971,PrakashChandra1971} to uniquely correspond to convex combinations of maximally mixed states in each photon-number subspace [alternative derivations were demonstrated by \textcite{Lehneretal1996,Soderholmetal2001}]:
\eq{
    \hat{\rho}_{\mathrm{isotropic}}=\sum_{N=0}^\infty \beta_N\frac{\hat{\mathds{1}}_N}{N+1},\quad \sum_{N=0}^\infty \beta_N=1,\quad \beta_N\geq 0.
    \label{eq:isotropic state}
} Here, the maximally mixed states are projections onto an $N$-photon subspace that can be written explicitly as
\eq{
    \hat{\mathds{1}}_N=\sum_{m=0}^N\ket{m}_\mathrm{R}\bra{m}\otimes\ket{N-m}_\mathrm{L}\bra{N-m}&=\sum_{m=0}^N\ket{m}_\mathrm{D}\bra{m}\otimes\ket{N-m}_\mathrm{A}\bra{N-m}\\
    &=\sum_{m=0}^N\ket{m}_\mathrm{H}\bra{m}\otimes\ket{N-m}_\mathrm{V}\bra{N-m}.
} The minimum distance between a given state and the isotropic states has been posited as a way of determining new quantum degrees of polarization \cite{Ghiuetal2010,Klimovetal2005,SanchezSotoetal2006,Bjorketal2010}.\footnote{This can give rise to new conceptions of maximally polarized states, different from what is discussed above \cite{SanchezSotoetal2007}; other quantum definitions of higher-order polarization have also been proposed \cite{SinghPrakash2013nonGaussian}.} Within a given photon-number subspace, these states have no remaining degrees of freedom, just like the classical description of unpolarized states. The only degrees of freedom in these states come from the probability distributions $\boldsymbol{\beta}$, which may be interpreted as the intensity distribution information to which classical polarization could, in theory, be privy. For example, the intensity for classical states may follow a Poisson distribution, taking the form
\eq{
    \hat{\rho}_{\mathrm{isotropic\,classical}}=\sum_{N=0}^\infty \frac{\left|\alpha\right|^{2N}\eu^{-\left|\alpha\right|^2}}{N!}\frac{\hat{\mathds{1}}_N}{N+1},
    \label{eq:isotropic Poisson}
} which could arise from the convex combination of the classically polarized states given in Eq. \eqref{eq:canonical coherent states any pol} averaged over all polarization directions.

The distinction between classically unpolarized states with $p=0$ and the completely isotropic states $\hat{\rho}_{\mathrm{isotropic}}$ can be summarized using the anticoherence concept uncovered by \textcite{Zimba2006}: classical unpolarization implies that $\expct{\left(\hat{\mathbf{S}}\cdot\mathbf{n}\right)}=0$ for all unit vectors $\mathbf{n}$, while quantum unpolarization implies that $\expct{\left(\hat{\mathbf{S}}\cdot\mathbf{n}\right)^k}$ is independent from $\mathbf{n}$ for larger integers $k$. States satisfying these constraints for the largest integers $k$ are now known as Kings of Quantumness \cite{Bjorketal2015,Bjorketal2015PRA} and have been explored numerically in many dimensions.

We present a final method for finding unpolarized states stemming from the classical definition. Given the decompositions of Eqs. \eqref{eq:Stokes unpol} and \eqref{eq:coherency unpol}, we are inspired to write
\eq{
    \hat{\rho}=p\hat{\rho}_{\mathrm{pol}}+(1-p)\hat{\rho}_{\mathrm{unpol}},
    \label{eq:quantum decomposition into classical}
} where $\hat{\rho}_{\mathrm{pol}}$ has degree of polarization $p=1$ and $\hat{\rho}_{\mathrm{unpol}}$ has $p=0$. Classically, we can thus always take any partially polarized state and subtract the polarized component to find the unpolarized component, such as through
\eq{
    \boldsymbol{\Psi}_{\mathrm{unpol}}=\frac{\boldsymbol{\Psi}-p\boldsymbol{\Psi}_{\mathrm{pol}}}{1-p}.
} Quantum mechanically, this is not always tenable. Making the same construction with quantum states, we have
\eq{
    \hat{\rho}_{\mathrm{unpol,\,candidate}}=\frac{\hat{\rho}-p\hat{\rho}_{\mathrm{pol}}}{1-p},
} but this is not unique, because there is no single unique $\hat{\rho}_{\mathrm{pol}}$ to subtract. Moreover, the resulting candidate is not always a quantum state, as it may fail to be positive. For example, given that the initial state may be pure, the subtracted state will always fail to be positive:
\eq{
    \hat{\rho}_{\mathrm{unpol,\,poor\,candidate}}=\frac{\ket{\psi}\bra{\psi}-p\hat{\rho}_{\mathrm{pol}}}{1-p}\leq 0.
} We thus find that this classical method for finding unpolarized states only \textit{sometimes} works in the quantum domain, requiring both a judicious choice of polarized component $\hat{\rho}_{\mathrm{pol}}$ and the verification that the resultant state is physically viable.

\subsection{Changes in Polarization}
What is the quantum perspective on classical polarization transformations? From the proceeding discussions, it will be clear that there are quantum transformations about which classical polarization is ignorant, while the quantum theory is fully cognisant of the classical transformations. We can analyze each of the classical transformations in turn.

\subsubsection{Quantum Transformations Underlying Jones Matrix Calculus}
We first inspect the quantum transformations that can be described by pure Jones matrix transformations, as in Eqs. \eqref{eq:coherency pure jones} and \eqref{eq:Mueller from pure Jones}. Classically, these are referred to as deterministic transformations, while we will see this notation to be at odds with some standard quantum nomenclature.

\label{sec:quantum rotations}
First, we consider polarization rotations. The Jones matrix given in Eq. \eqref{eq:Jones rotation} directly corresponds to its quantum counterpart, where a rotation operator is defined by
\eq{
    \hat{R}\left(\Theta,\mathbf{n}\right)=\exp\left(-\iu\Theta\mathbf{n}\cdot\hat{\mathbf{S}}\right).
    \label{eq:quantum rotation}
} When a quantum state undergoes a polarization rotation
\eq{
    \hat{\rho}^{(\mathrm{in})}\to \hat{\rho}^{(\mathrm{out})}=\hat{R}\hat{\rho}^{(\mathrm{in})}\hat{R}^\dagger,
} the Stokes operators transform as
\eq{
    \hat{\mathbf{S}}^{(\mathrm{in})}\to \hat{\mathbf{S}}^{(\mathrm{out})}=\hat{R}\hat{\mathbf{S}}^{(\mathrm{in})}\hat{R}^\dagger=\mathbf{R}\hat{\mathbf{S}},
} where $\mathbf{R}$ is the $3\times 3$ rotation matrix found in Eq. \eqref{eq:Mueller rotation}.
This type of transformation is unitary, is known as an SU(2) rotation, and leaves $\hat{S}_0$ unchanged, thereby allowing the Stokes \textit{operators} to transform in the same way as the Stokes parameters, through
\eq{
    \hat{S}_\mu\to\sum_{\nu=0}^3 M_{\mu \nu}\hat{S}_\nu.
} In fact, because we seek descriptions of polarization transformations that remain valid regardless of the input state, it will remain a generic feature that the Stokes operators transform via the Mueller matrices describing the transformations of the associated Stokes parameters.

When acting on creation and annihilation operators, the rotation operations enact
\eq{
    \hat{R}\hat{A}\hat{R}^\dagger=\mathbf{J}_{\mathrm{rot}}\hat{A},
}for the quantized Jones vector [cf. Eq. \eqref{eq:quantized Jones vector}]
\eq{
    \hat{A}=\begin{pmatrix}
        \ha\\\hb
    \end{pmatrix}.
} 
This is what guarantees that the Majorana constellation rotates rigidly under a polarization rotation, as the creation operators $\hat{a}^\dagger_\Omega$ have their angular coordinates rotate together through
\eq{
    \hat{R}\ket{\psi^{(N)}}\propto\hat{R}\prod_{k=1}^N \hat{a}^\dagger_{\Omega_k}\vac=\left(\prod_{k=1}^N \hat{R}\hat{a}^\dagger_{\Omega_k}\hat{R}^\dagger\right) \hat{R}\vac=\left(\prod_{k=1}^N \hat{R}\hat{a}^\dagger_{\Omega_k}\hat{R}^\dagger\right) \vac.
}
In addition, that the Stokes operators themselves transform in the same way as the Stokes parameters means that higher-order moments such as $\expct{\hat{S}_i\hat{S}_j}$ also transform as expected classically under polarization rotations, albeit under the assumption that the classical values for operator correlations are already correctly given by the quantum expectation values.
These facts make the quantum rotation transformations very similar to their classical counterparts, explaining the true origin of classical polarization rotations.

Arbitrary Jones matrices acting on $A$ cannot, in contrast to rotations, simply act on the quantized vector $\hat{A}$. The diattenuation transformation of Eq. \eqref{eq:Jones diattenuation} applied to quantized fields through
\eq{
    \ha\to\sqrt{q}\ha,\quad \hb\to\sqrt{r}\hb,
}
for example, is not a trace-preserving quantum channel and does not preserve the bosonic commutation relations of the two modes. To make these transformations preserve unitarity, an auxiliary, possibly fictious, extra pair of modes annihilated by some bosonic operators $\hat{v}_1$ and $\hat{v}_2$ must be introduced, to create transformations of the form
\eq{
    \ha\to\sqrt{q}\ha-\sqrt{1-q}\hat{v}_1,\quad \hb\to\sqrt{r}\hb-\sqrt{1-r}\hat{v}_2.
    \label{eq:quantum diattenuation two input outputs}
} Then, ignoring the auxiliary modes leads to an \textit{effective} transformation that looks like that of a diattenuation of the two modes $a$ and $b$ in which the underlying physical transformation implies that some photons from those two modes were transferred to auxiliary modes.

We can consider an attenuation transformation to be a rotation between some polarization mode $a$ and another mode $v_1$ initially in its vacuum state. Physically, this is also equivalent to introducing a beam splitter that intermixes modes $a$ and $v_1$ and then ignores the latter mode. The transmission probability of the beam splitter or the effective transmission probability of the fictitious beam splitter is exactly equal to the attenuation coefficient $q$. Notably, since the effect of sequential attenuations can be collated into that of a single attenuation by a larger factor $q=q_1\cdots q_n$, we only need a single rotation matrix with a single auxiliary mode to describe attenuation from a quantum standpoint.

A quantum state undergoing attenuation in mode $a$ has the quantum channel
\eq{
    \hat{\rho}\to\sum_l \hat{K}_l \hat{\rho}\hat{K}_l,\quad \sum_l \hat{K}_l^\dagger\hat{K}_l^\dagger=\hat{\mathds{1}}\label{eq:quantum channel Kraus},
} with Kraus operators given by \cite{Goldberg2021thesis}
\eq{
    \hat{K}_l=\sum_{m=0}^\infty\sqrt{\binom{m+l}{m}q^l\left(1-q\right)^m}\ket{m}_a\bra{m+l}.
} The Stokes operators for such a process transform, in turn, as
\eq{
    \hat{S}_\mu\to\sum_l \hat{K}_l^\dagger \hat{S}_\mu\hat{K}_l
    \label{eq:quantum channel Kraus on Stokes}.
} This can be combined with an attenuation in the second mode to yield the classical transformations of Eq. \eqref{eq:Jones diattenuation}.

\begin{figure*}
    \centering
    \includegraphics[width=\textwidth]{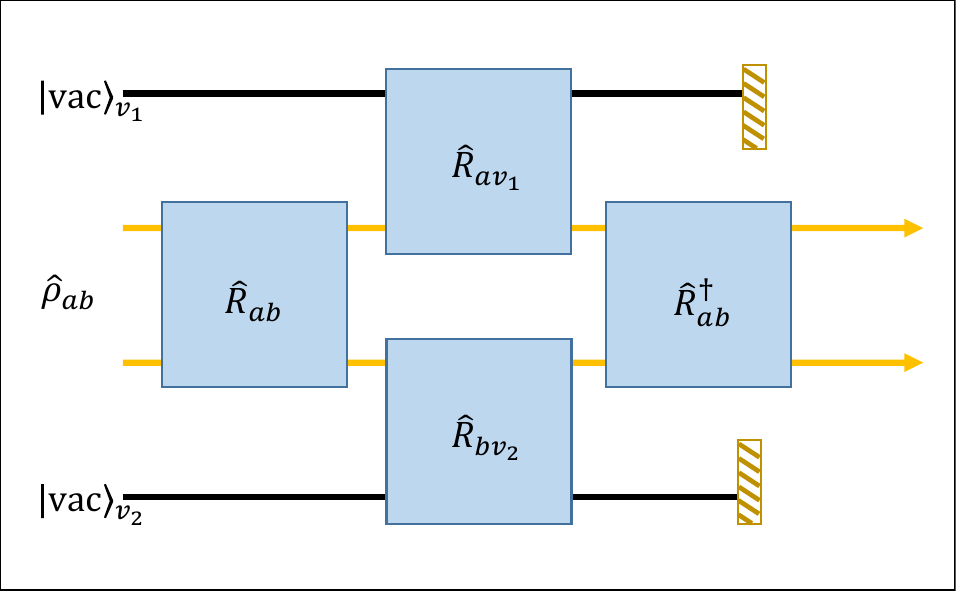}
    \caption{Quantum circuit diagram underlying classical polarization diattenuations. The input polarization state $\hat{\rho}_{ab}$ is first rotated to a desired configuration, the two modes are each attenuated via a rotation with their auxiliary vacuum modes $v_1$ and $v_2$, the vacuum modes are ignored, the polarization modes are rotated back to the original configuration, and the output two arrows correspond to the diattenuated polarization state.}
    \label{fig:diattenuation general}
\end{figure*}

As seen in the classical picture of Eq. \eqref{eq:Mueller diattenuation general}, the most general diattenuation transformation has four free parameters: two govern the strengths of the attenuations and two govern the pair of orthogonal polarization modes being attenuated. These can all be accounted for using rotation operations: two polarization rotations enable the basis changes to find the modes to be attenuated and two rotations into vacuum modes perform the attenuations. This most general diattenuation transformation is schematized in Fig. \ref{fig:diattenuation general}. When the vacuum modes are ignored, it is as if the quantized Jones vector undergoes a classical diattenuation transformation
\eq{
    \hat{A}\to\mathbf{J}\hat{A}.
} 

The marked difference between polarization rotations and diattenuations is that the former is a unitary transformation on the polarization state and the latter is not, instead given by a quantum channel of the form of Eq. \eqref{eq:quantum channel Kraus} with more than one Kraus operator $\hat{K}_l$. Regrettably, the quantum theory refers to unitary channels as deterministic and nonunitary channels as nondeterministic, even though both transformations are classically referred to as deterministic from a polarization standpoint. 

It is noteworthy that all of transformations governed by a single Jones matrix, classically referred to as deterministic, can be considered to conserve photon number in some enlarged Hilbert space. We will see this to change when inspecting classically nondeterministic polarization transformations.

\subsubsection{Quantum Transformations Underlying Mueller Matrix Calculus}
It is straightforward to generalize the quantum transformations underlying ``deterministic'' polarization transformations to those underlying transformations of the form of Eqs. \eqref{eq:coherency multiple Jones with weights} and \eqref{eq:Mueller from Jones with weights} with more than one nonzero weight $\lambda_i$. Under the standard assumption that the classical weights can only be positive, the quantum transformations immediately follow as probabilistic mixtures of the underlying quantum transformations. These can be seen as the nondeterministic transformations
\eq{
    \hat{\rho}\to\sum_i\lambda_i\sum_l \hat{K}_l^{(i)}\hat{\rho}\hat{K}_l^{(i)\dagger} \quad\Rightarrow \quad \mathbf{M}=\sum_i \lambda_i \mathbf{M}^{(i)},
} where each Mueller matrix $\mathbf{M}^{(i)}$ arises from a deterministic polarization transformation with Kraus operators $\hat{K}_l^{(i)}$. The new transformations are now governed by the larger set of Kraus operators $\sqrt{\lambda_i}\hat{K}_l^{(i)}$, indexed by both $i$ and $l$, which can also be used in Eq. \eqref{eq:quantum channel Kraus on Stokes} to describe the most general quantum mechanical transformation on the Stokes operators that leads to Mueller matrix transformations of the Stokes parameters. 

Many examples serve to tease apart the nuances of quantum polarization transformations. For example, consider a classical transformation whereby a state has an equal probability of undergoing one of two rotations parametrized as $\mathbf{R}_i=\mathbf{R}\left(\Theta_i,\mathbf{n}_i\right)$. Quantum mechanically, this may arise by a unitary operation between a polarization state $\ket{\psi}_{ab}$ and an auxiliary mode initially in its vacuum state as
\eq{
    \ket{\psi}_{ab}\otimes \ket{0}_v\to\frac{\hat{R}_1\ket{\psi}_{ab}\otimes\ket{1}_v+\hat{R}_2\ket{\psi}_{ab}\otimes\ket{2}_v}{\sqrt{2}}. 
} Since the amount of rotation becomes entangled with the auxiliary mode, which may have gone a photon-number \textit{non}conserving operation, ignoring the vacuum mode leads to the effective polarization transformation
\eq{
    \hat{\rho}_{ab}\to\frac{1}{2}\hat{R}_1\hat{\rho}\hat{R}_1^\dagger+\frac{1}{2}\hat{R}_2\hat{\rho}\hat{R}_2^\dagger.
}
This is equivalent to a transformation with the pair of Kraus operators $\hat{K}_i=\hat{R}_i/\sqrt{2}$ that manifestly satisfy the normalization requirement of Eq. \eqref{eq:quantum channel Kraus}.
Even though this operation is unitary in a larger Hilbert space, it is markedly different from the simple rotation transformations that enact diattenuations in larger Hilbert spaces.

\subsubsection{Speculative Constraints on Polarization Changes}
Quantum channels acting on a quantum state must be completely positive. A general further assumption is that they preserve the trace of the quantum state, so as to preserve total probability. How do these considerations, encompassed by Eq. \eqref{eq:quantum channel Kraus}, constrain the possible Mueller matrix transformations of Eq. \eqref{eq:Mueller definition Stokes}?

We do not yet have an answer to this question. It is clear from the above sections that all classical transformations falling under the assumptions of Section \ref{sec:classical constraints} can be reproduced by the quantum theory. Can we justify these assumptions by assuming only quantum theory? Can we circumvent these assumptions using quantum theory? As mentioned before, these questions have been touched on previously by \textcite{AhnertPayne2005,Aielloetal2007,Sudhaetal2008,Simonetal2010,GamelJames2011}, but we believe them to remain unanswered.

The linearity assumption is easiest to break, but that can be broken using both classical and quantum perspectives. Namely, many transformations ascribing to the form of Eq. \eqref{eq:quantum channel Kraus on Stokes} do not lead to linear transformations among the Stokes parameters; an easy example is a unitary operation corresponding to a nonlinear Hamiltonian, such as
\eq{
    \hat{U}=\exp\left(-\iu \chi \hat{S}_3^2\right).
} Similarly, not all classical transformations are linear, as with light experiencing the Kerr effect, so we should not expect every operation in the universe to produce a transformation with a Mueller matrix as in Eq. \eqref{eq:Mueller definition Stokes}. In fact, nonlinear polarimetry is a field unto itself that merits its own attention \cite{Bazhenovetal1994,BrasseletZyss2007,Samimetal2016PRA,Samimetal16JOSAB,KrouglovBarzda2019}.
These considerations let us refine our current questions to ask whether quantum considerations affect the constraints of classical \textit{linear} polarization transformations that are simply governed by Mueller matrices as in Eq. \eqref{eq:Mueller definition Stokes}.

The most tantalizing question is whether quantum transformations alone can be used to restrict the weights in Eqs. \eqref{eq:coherency multiple Jones with weights} and \eqref{eq:Mueller from Jones with weights} to be positive, as conjectured by \textcite{Goldberg2020}. Quantum transformations restricted to the single-photon subspace automatically necessitate the positivity assumption \cite{GamelJames2011}, so our conjecture would be proven if one could prove that all quantum transformations that enforce linear transformations among the Stokes parameters must necessarily take single-photon states to single-photon states, but we have already seen that diattenuations do not maintain a single photon-number subspace. Similarly, our conjecture could be proven if one could show that the only quantum transformations that enable linear transformations among the extended Stokes parameters of Eq. \eqref{eq:two point Stokes parameters} are those that enact linear transformations among the regular Stokes parameters. It would be nice to use the SU(4)--O$^+$(6) homomorphism discussed in the classical context of \textcite{Cloude1986} to attack this problem from a quantum standpoint, but not all Mueller matrices are unitary, so there indeed remains work to be done.

We can also continue our classical discussion of the restriction on Mueller matrices to satisfy the transmittance and reverse transmittance conditions. It is evident that two Kraus operators $\hat{K}_1=\mathbf{J}_1$ and $\hat{K}_2=\mathbf{J}_2$ expressed in the single-photon basis as in Eq. \eqref{eq:lossless polarizer Jones} lead to the transformation
\eq{
    \hat{\rho}_{\mathrm{single\,photon}}\to \sum_{l=1}^2\hat{K}_l\hat{\rho}_{\mathrm{single\,photon}}\hat{K}_l^\dagger=\ket{1}_{\mathrm{R}}\bra{1}\otimes\ket{0}_{\mathrm{L}}\bra{0}.
    \label{eq:Kraus to single photon}
} Somehow, there exists a quantum mechanical transformation that leads to the Mueller matrix of Eq. \eqref{eq:lossless polarizer Mueller}, when restricted to act on single-photon input states, that violates the reverse transmittance condition. Have we uncovered a contradiction between the theories?

As with all paradoxes in the technical sense of the word, a resolution awaits. The creation of such an input-agnostic polarization transformer from linear optical devices requires postselection to enable a lossless polarizer \cite{Zhangetal2021arxiv}; in actuality, some light is always lost by such a polarizer. Such a transformation with Kraus operators enacting Eq. \eqref{eq:Kraus to single photon}
certainly exists, because it is always possible to create a quantum channel on a finite-dimensional Hilbert space that takes arbitrary input states to a fixed output state \cite{Wuetal2007}, but it raises concerns that we presently address: What about infinite-dimensional input states? Is this transformation really linear?

One can concoct a quantum channel that transforms arbitrary input states into a given pure output state $\ket{\psi_{\mathrm{target}}}$ by amassing an infinite set of Kraus operators that transform a complete set of basis states into the desired final state:
\eq{
    \hat{K}_{m,n}=\ket{\psi_{\mathrm{target}}}\bra{m,n},\qquad \forall\,m,n\in \left(0,1,\cdots,\infty\right).
    \label{eq:Kraus to pure state}
} These Kraus operators readily satisfy the normalization constraint of Eq. \eqref{eq:quantum channel Kraus} and so represent a viable physical transformation; similar results can be obtained if the target state is mixed \cite{Wuetal2007}. However, such a channel does not enact a linear transformation among the Stokes parameters, as it creates output states with the same total intensity regardless of the input intensity. We can modify this scheme to create an ideal lossless polarizer that polarizes all input light into the direction $\Omega$ by using the infinite set of Kraus operators
\eq{
    \hat{K}_{m,n}=\ket{\Omega^{(m+n)}}\bra{m,n},\qquad \forall\,m,n\in \left(0,1,\cdots,\infty\right),
    \label{eq:Kraus to polarized state}
} which again satisfies all of the requirements of a quantum channel. This appears to act linearly, as it directly enacts a Mueller matrix with the form of Eq. \eqref{eq:lossless polarizer Mueller}:
\eq{
    \mathbf{M}=\begin{pmatrix}
        1&\mathbf{0}^\top\\
        \mathbf{n}_\Omega&\mathbf{0}\mathbf{0}^\top
    \end{pmatrix}
} (as a reminder, the outer product $\mathbf{0}\mathbf{0}^\top$ equals a $3\times 3$ matrix of zeroes). Quantum theory here contradicts the standard assumptions of classical polarization theory.

For many reasons, one should not expect transformations with Kraus operators such as those in Eqs. \eqref{eq:Kraus to pure state} and \eqref{eq:Kraus to polarized state} to ever be feasible in practice. The Kraus operators can be physically interpreted as a measure-and-prepare operation that checks what the input state is and outputs some state depending on that input. Currently, Kraus operators as in Eq. \eqref{eq:Kraus to polarized state} must measure the intensity of the input state and coherently output a perfectly polarized state with that same intensity in order to act like a lossless polarizer. Such a measurement process alone is unlikely to be linear, but an even greater challenge is to create a physical setup that can enact this transformation for arbitrary input intensities (and, likewise, arbitrary superpositions and convex combinations of intensities).
Additionally, when considering these Kraus operators as arising from a unitary operation in an enlarged Hilbert space, the unitary must necessarily create a \textit{different} output state when the initial state of the auxiliary system $v$ is different. For example, the single-photon transformation from Eq. \eqref{eq:Kraus to single photon} must be constructed from a unitary of the form
\eq{
    \hat{U}&=\ket{R}\bra{R}\otimes\ket{0}_v\bra{0}+\ket{R}\bra{L}\otimes\ket{1}_v\bra{0}+
    \ket{L}\bra{R}\otimes\ket{0}_v\bra{1}+\ket{L}\bra{L}\otimes\ket{1}_v\bra{1},\, \mathrm{or}\\
    \hat{U}&=\ket{R}\bra{R}\otimes\ket{0}_v\bra{0}+\ket{R}\bra{L}\otimes\ket{1}_v\bra{0}+
    \ket{L}\bra{L}\otimes\ket{0}_v\bra{1}+\ket{L}\bra{R}\otimes\ket{1}_v\bra{1},
} up to a relabelling of the auxiliary mode. This performs the correct transformation on $\hat{\rho}_{\mathrm{single\,photon}}\otimes\ket{0}_v\bra{0}$, sending it to a right-handed circularly polarized state, while doing the opposite to $\hat{\rho}_{\mathrm{single\,photon}}\otimes\ket{1}_v\bra{1}$, sending it to a left-handed circularly polarized state, so the device would have to be reset each time to ensure the proper input state in the auxiliary system.
The infeasibility of such operations is then a quantum justification for the polarization transmittance conditions that are classically assumed.
Future work could certainly bolster this connection, which we have only here begun to undertake.

\section{Quantum Polarimetry from the Perspective of Quantum Estimation Theory}
\label{sec:quantum polarimetry estimation}
The crux of polarimetry is the ability to measure polarization and its changes \textit{well}. There has been significant research showing that techniques from or inspired by quantum theory can be used to outperform their classical counterparts for particular tasks in sensing and metrology \cite{Caves1981,Dowling1998,Mitchelletal2004,Giovannettietal2004,Berryetal2009,LIGO2011,Humphreysetal2013,Tayloretal2013,Tsangetal2016,Liuetal2020}, so quantum polarimetry offers the potential for similar quantum advantages. Polarimetry tasks have been explored directly from a quantum perspective in a variety of guises that we presently review.

\subsection{Quantum Fisher Information}
A central quantity in quantum sensing applications is the quantum Fisher information matrix (QFIM), which characterizes how sensitive a given probe state is to changes in the parameters being measured. The QFIM differs from its classical counterpart in that it is optimized over all possible measurement strategies such that it depends only on the probe state and the parameters being measured. This facilitates a direct search for the most sensitive probe states, which can be used to produce estimates of the underlying parameters with the lowest possible uncertainties.

We start with a few properties of QFIMs \cite{Matsumoto2002,Paris2009,TothApellaniz2014,Szczykulskaetal2016,Braunetal2018,Liuetal2019,SidhuKok2020,Albarellietal2020,Polinoetal2020,DemkowiczDobrzanskietal2020,Liuetal2021accepted}. Given a series of of parameters to be estimated $\btheta$, we wish to find as small as possible of a covariance matrix for an estimator $\tilde{\btheta}$
\eq{
    \Cov\left(\tilde{\btheta}\right)=\expct{\left(\tilde{\btheta}-\btheta\right)\left(\tilde{\btheta}-\btheta\right)^\top},
} where we have assumed unbiased estimators\footnote{Biased estimators have indeed been investigated in the context of quantum metrology but may be harder to implement as they depend on the underlying parameters $\btheta$ \cite{Motkaetal2016,BonsmaFisheretal2019}.} such that
\eq{
    \expct{\tilde{\btheta}}=\btheta.
} Then, the pivotal result from quantum estimation theory is that the covariance matrix for a measurement repeated $\nu\gg 1$ times is bounded from below by the inverse of the QFIM through the quantum Cram\'er-Rao bound (qCRB):
\eq{
    \Cov\left(\tilde{\btheta}\right) \geq \nu^{-1} \mathbf{Q}\left(\btheta;\hat{\rho}_{\btheta}\right)^{-1}.
    \label{eq:qCRB}
} The QFIM is computed from the evolved probe state $\hat{\rho}_{\btheta}$ as
\eq{
    Q_{i j}\left(\btheta;\hat{\rho}_{\btheta}\right)=\Tr\left(\hat{\rho}_{\btheta}\frac{\left\{\hat{L}_{\theta_i},\hat{L}_{\theta_j}\right\}}{2}\right),
} where the symmetric logarithmic derivative operators are implicitly defined through
\eq{
    \frac{\partial \hat{\rho}_{\btheta}}{\partial \theta_i}=\frac{\left\{\hat{L}_{\theta_i},\hat{\rho}_{\btheta}\right\}}{2}
} and we are using the anticommutator $\left\{A,B\right\}=AB+BA$. For pure probe states with unitary generators
\eq{
    \frac{\partial \hat{\rho}_{\btheta}}{\partial \theta_i}=\iu\left[\hat{\rho}_{\btheta},\hat{G}_i\right]
} the QFIM takes the particularly simple form
\eq{
    Q_{i j}\left(\btheta;\hat{\rho}_{\btheta}\right)=4\Cov_{\hat{\rho}_{\btheta}}\left(\hat{G}_i,\hat{G}_j\right),
} where we are using the symmetrized covariance $\Cov_{\hat{\rho}}\left(A,B\right)=\expct{\left\{A,B\right\}}/2-\expct{A}\expct{B}$ that takes expectation values with respect to state $\hat{\rho}$. When there is a single parameter $\theta$ to be estimated, the qCRB of Eq. \eqref{eq:qCRB} is always saturable. For multiparameter estimation, the bound is saturable when, for all $i$ and $j$,
\eq{
    \Tr\left(\hat{\rho}_{\btheta}\left[\hat{L}_{\theta_i},\hat{L}_{\theta_j}\right]\right)=0.
    \label{eq:commutativity condition}
} Many other details about quantum estimation can be found in the above-mentioned reviews and we, further, draw attention to recent work investigating situations in which the QFIM in Eq. \eqref{eq:qCRB} cannot be inverted or changes discontinuously \cite{Safranek2017,GoldbergJames2018Euler,ZhouJiang2019arxiv,Sevesoetal2019,YeLual2021arxiv,Goldbergetal2021singularaccepted}.

\subsection{Rotation Measurements}
\subsubsection{Phase Estimation}
One main quantum-enhanced measurement protocol is mathematically equivalent to a subset of rotation measurements: phase estimation. Many physical processes amount to estimating a single parameter $\theta$ from a unitary of the form $\hat{U}=\exp\left(-\iu\theta\hat{G}\right)$. In the language of polarization, this arises when trying to estimate the angle of a polarization rotation about a known axis such as $\mathbf{n}=\mathbf{e}_3$:
\eq{
    \hat{U}(\theta)=\hat{R}\left(\theta,\mathbf{z}\right)=\exp\left(-\iu\theta\hat{S}_3\right).
} Using a classical state for such an estimation problem leads to the evolution
\eq{
    \ket{\psi_{\mathrm{classical}}(\theta)}=\hat{U}(\theta)\left(\ket{\alpha}_{\mathrm{R}}\otimes\ket{0}_{\mathrm{L}}\right)=\ket{\alpha\eu^{-\iu\frac{\theta}{2}}}_{\mathrm{R}}\otimes\ket{0}_{\mathrm{L}}.
} We can compute the QFIM as a function of the average photon number (i.e., energy) of the input light
\eq{
    H=\expct{\hat{N}}
} to find the ``shot-noise-scaling'' QFIM
\eq{
    Q\left(\theta;\ket{\psi_{\mathrm{classical}}}\right)=H.
} A NOON state such as that of Eq. \eqref{eq:NOON} with $\Theta=0$ is much more sensitive, evolving as
\eq{
    \ket{\psi_{\mathrm{NOON}}(\theta)}=\hat{U}(\theta)\frac{\ket{N,0}+\ket{0,N}}{\sqrt{2}}=\frac{\eu^{-\iu\frac{\theta}{2}}\ket{N,0}+\eu^{\iu\frac{\theta}{2}}\ket{0,N}}{\sqrt{2}}.
} This leads to a ``Heisenberg-scaling'' QFIM
\eq{
    Q\left(\theta;\ket{\psi_{\mathrm{NOON}}}\right)=H^2,
} allowing for much more sensitive measurements than the classical scheme with a commensurate amount of input energy.

This advantage in phase estimation is readily extendible to simultaneously estimating multiple phases \cite{Humphreysetal2013}. There, an additional advantage is seen versus the sequential estimation of each of the parameters. Caveats exist in terms of the availability of a reference mode \cite{JarzynaDemkowiczDobrzanski2012,Goldbergetal2020multiphase} and the number of required measurements \cite{GoreckiDemkowiczDobrzanski2021arxiv}, but we cannot consider generalized polarimetry to be a multiphase estimation protocol because the polarimetric parameters are not all mutually independent. Instead, we continue to investigate quantum-enhanced polarimetry along the standard decompositions of polarization transformations.

\subsubsection{Rotations about Unknown Axes}
A more intricate estimation problem is that of estimating both the rotation angle and rotation axis of an unknown rotation. This has been done using a variety of parametrizations for the three parameters of a rotation \cite{BaumgratzDatta2016,GoldbergJames2018Euler,Frieletal2020arxiv,Houetal2020,Goldbergetal2021rotationspublished}, which can all be unified from a geometrical perspective \cite{Goldbergetal2021intrinsic}.

No matter the parametrization of the triad of rotation parameters $\btheta$, rotation operations are unitary, per Eq. \eqref{eq:quantum rotation}. This facilitates three unitary generators of the transformation
\eq{
    \hat{G}_i=\iu\frac{\partial \hat{R}\left(\btheta\right)}{\partial \theta_i}\hat{R}^\dagger\left(\btheta\right),
} one for each parameter $\theta_i$. Since these can be shown to be comprised from linear combinations of Stokes operators \cite{Goldbergetal2021rotationspublished,Goldbergetal2021intrinsic}, in the form of
\eq{
    \hat{G}_i=\mathbf{g}\cdot\mathbf{S},
}they can readily be computed using any representation of SU(2), which is especially straightforward using Pauli matrices. The QFIM then takes the form
\eq{
    \mathbf{Q}\left(\btheta;\psi\right)=4\mathbf{G}\left(\btheta\right)^\top\mathbf{C}\left(\psi_{\btheta}\right)\mathbf{G}\left(\btheta\right), \quad \mathbf{G}=\begin{pmatrix}
        \mathbf{g}_1 &\mathbf{g_2} &\mathbf{g}_3
    \end{pmatrix}, \quad \mathbf{C}_{ij}\left(\psi\right)=\Cov_{\ket{\psi}\bra{\psi}}\left(\hat{S}_i,\hat{S}_j\right),
} which, by rotating the covariance matrix to that of the unrotated state via $\mathbf{C}\left(\psi_{\btheta}\right)=\mathbf{R}\left(\btheta\right)^\top\mathbf{C}\left(\psi\right)\mathbf{R}\left(\btheta\right)$, allows us to fully separate the parametric dependence of the QFIM from its state dependence:
\eq{
    \mathbf{Q}\left(\btheta;\psi\right)=4\pmb{G}\left(\btheta\right)^\top\mathbf{C}\left(\psi\right)\pmb{G}\left(\btheta\right), \quad \pmb{G}=\mathbf{R}\mathbf{G}.
} This allows one to maximize the ``sensitivity covariance matrix'' over all probe states $\ket{\psi}$ without having to worry about the absolute values of the parameters or the parametrization being considered. 

The qCRB of Eq. \eqref{eq:qCRB} is a matrix bound, which is hard to uniquely optimize. A scalar bound can be found by using a positive-definite weight matrix $\mathbf{W}$ to produce the lower bound for a certain linear combination of the parameters' variances and covariances
\eq{
    \Tr\left[\mathbf{W}\Cov\left(\tilde{\btheta},\tilde{\btheta}^\top\right)\right] \geq\Tr\left[\mathbf{W} \mathbf{Q}\left(\btheta;\hat{\rho}_{\btheta}\right)^{-1}\right].
} When the weight matrix is chosen to be the metric tensor for SU(2), $\mathbf{W}=\boldsymbol{\eta}$, all of the matrices $\pmb{G}$ cancel and we are left with the scalar qCRB for the weighted mean-square error wMSE \cite{Goldbergetal2021intrinsic}
\eq{
    \mathrm{wMSE}(\tilde{\btheta})=\Tr\left[\boldsymbol{\eta}\Cov\left(\tilde{\btheta},\tilde{\btheta}^\top\right)\right] \geq\Tr\left[\mathbf{C}\left(\psi\right)^{-1}\right].
} This can then be uniquely optimized to find the most sensitive probe states for simultaneously estimating all three parameters of a rotation.

Probe states with the smallest values of $\mathrm{wMSE}(\tilde{\btheta})$ for a given average energy $H$ have a few properties \cite{GoldbergJames2018Euler,Goldbergetal2020extremal,Goldbergetal2021rotationspublished,Goldbergetal2021intrinsic}:
\begin{itemize}
    \item They are pure states
    \item They have definite total spin $S_0=H/2$.
    \item They are classically unpolarized ($p=0$), meaning that the Stokes operators have vanishing expectation value $\hat{\mathbf{S}}=\mathbf{0}$.
    \item They are unpolarized to second order, making them anticoherent states or Kings of Quantumness to second order, with isotropic sensitivity covariance matrices $\mathbf{C}\left(\psi\right)\propto \mathds{1}$.
\end{itemize}
The tetrahedral state given in Fig. \ref{fig:Majorana examples} is an example of such an ideal probe state, with other examples also having highly symmetric Majorana constellations.
They achieve the Heisenberg-scaling lower bound
\eq{
    \mathrm{wMSE}(\tilde{\btheta})\geq 4\frac{9}{H\left(H+1\right)},
} which can be saturated by an optimal measurement scheme \cite{Goldbergetal2021rotationspublished}.
NOON states, by contrast, only have symmetries about a single axis in their Majorana constellations, so they are no longer superior in rotation estimation, instead achieving only shot-noise scaling
\eq{
    \mathrm{wMSE}(\tilde{\btheta})\geq 4\frac{2H+1}{H^2}.
    \label{eq:NOON state MSE}
} Such ideal second-order unpolarized states have been generated using light's orbital angular momentum degree of freedom \cite{Bouchardetal2017} and work is underway to do likewise in the polarization degrees of freedom.

The qCRB is saturable for the simultaneous estimation of all three rotation parameters for all pure unpolarized states. This is in stark contrast to classical light, which we saw in Eqs. \eqref{eq:uncertainty limit classical state 1} and \eqref{eq:uncertainty limit classical state 2} to obey stricter inequalities due to them violating the commutativity condition of Eq. \eqref{eq:commutativity condition} \cite{Jarzyna2021arxiv}. The nature of the optimal detection scheme for a given probe state thereby depends on the polarization properties of the probe, with the ultimate limit being achieved by second-order unpolarized states of light.

\subsection{Diattenuation Measurements}
It is also possible to obtain quantum enhancements in diattenuation measurements. These are physically equivalent to transmission and absorption measurements, which have been shown to benefit from quantum enhancements \cite{JakemanRarity1986,Heidmannetal1987,Hayatetal1999,MonrasParis2007,Brambillaetal2008,Jietal2008,Adessoetal2009,Alipouretal2014,Crowleyetal2014,Medaetal2017,Nair2018,Loseroetal2018,Ioannouetal2020arxiv}; however, these enhancements are modest, as they do not allow for improved scaling with probe energy beyond the shot-noise limit.

\subsubsection{Attenuation}
As described above, attenuation enacts the input-output relation 
\eq{
    \ha\to\sqrt{q}\ha+\sqrt{1-q}\hat{v},
} where mode $a$ is being attenuated and mode $v$ is initially in its vacuum state. This evolution also governs absorption, reflection, and transmission measurements, all of which seek to estimate $q$ or a related parameter. A canonical coherent state evolves as
\eq{
    \ket{\alpha}\to\ket{\sqrt{q}\alpha}=\eu^{\left(1-q\right)\left|\alpha\right|^2/2}q^{\had\ha/2}\ket{\alpha},
} yielding the QFIM
\eq{
    Q(q;\ket{\alpha})=\frac{H}{q},\quad H=\left|\alpha\right|^2.
} The optimal quantum strategy, by contrast, uses Fock states $\ket{N}$ that evolve as
\eq{
    \ket{N}\bra{N}\to\sum_{k=0}^N \binom{N}{k} q^k \left(1-q\right)^{N-k} \ket{k}\bra{k}.
} The coefficients in this probability distribution take the same form as for SU(2)-coherent states in Eq. \eqref{eq:coherent state amplitudes} because attenuations are akin to rotations with an ignored auxiliary mode. Fock states have the improved QFIM
\eq{
    Q(q;\ket{N})=\frac{H}{q\left(1-q\right)},
} which is especially helpful in the limit of small losses $q\approx 1$. No enhanced scaling with $H$ is possible, so the improvements are not as dramatic as for enhanced rotation sensing and can equally be achieved with $N$ single-photon states or a single $N$-photon state. This has been demonstrated with single photons heralded using spontaneous parametric downconversion \cite{YabushitaKobayashi2004,Bridaetal2010,Moreauetal2017,Samantarayetal2017,SabinesChesterkingetal2017,Whittakeretal2017,Yoonetal2020}.

Practically speaking, it is often easier to create two-mode squeezed vacuum (TMSV) states of the form
\eq{
    \ket{\psi_{\mathrm{TMSV}}}=\frac{1}{\cosh \zeta}\sum_{N=0}^\infty \left(-\eu^{\iu\varphi}\tanh \zeta\right)^N\ket{N,N},
}with average energy $H=2\sinh^2\zeta$, than Fock states $\ket{N}$. In fact, this is what is generated via spontaneous parametric downconversion when no heralding is performed. These states evolve as
\eq{
    \ket{\psi_{\mathrm{TMSV}}}\bra{\psi_{\mathrm{TMSV}}}\to\sum_{m=0}^\infty\left(1-q\right)^m\ket{\psi_m}\bra{\psi_m}
} for the mutually orthogonal, unnormalized states
\eq{
    \ket{\psi_m}=\frac{1}{\cosh\zeta}\sum_{N=0}^\infty\sqrt{\binom{N+m}{m}} \left(-\eu^{\iu\varphi}\sqrt{q}\tanh\zeta\right)^N\ket{N,N+m}.
} 
TMSV states have the same QFIM as Fock states to lowest order in $H$, making them similarly useful in the limit of small $q$ and small $H$ for providing quantum advantages, which has indeed been demonstrated \cite{Tapsteretal1991,SoutoRibeiroetal1997,DAuriaetal2006,Shietal2020,Atkinsonetal2021}.


\subsubsection{Diattenuation}
The estimation of two attenuation factors for orthogonal modes can simply consist of two parallel single attenuation measurements. Seeing that a Fock state $\ket{N}$ is optimal for sensing a single attenuation parameter $q$, a pair of Fock states is ideal for measuring a pair of attenutation parameters:
\eq{
    \mathbf{Q}\left(q,r;\ket{N_1,N_2}\right)=H\begin{pmatrix}
        \frac{h_1}{q\left(1-q\right)}&0\\ 0&\frac{h_2}{r\left(1-r\right)}
    \end{pmatrix},
} where $h_i=N_i/\left(N_1+N_2\right)$ is the energy fraction in each mode. This is a special sort of quantum probe state that is only partially polarized with degree of polarization $p=\left|h_1-h_2\right|$, in contrast to a fully polarized classical probe state that achieves
\eq{
    \mathbf{Q}\left(q,r;\ket{\alpha_\Omega}\right)=H\begin{pmatrix}
        \frac{\cos^2\Theta}{q}&0\\ 0&\frac{\sin^2\Theta}{r}
    \end{pmatrix}.
}

One could, in theory, use a pair of TMSV states to measure two attenuation parameters in parallel. This, however, presents physical challenges because the two polarization modes being attenuated are in the same spatial mode, making it difficult to entangle each polarization mode with a different external mode. In this multiparameter context, it is possible to directly use a single TMSV state to measure diattenuation, where the two modes are the polarization modes being attenuated. We have not found this idea directly discussed in the literature, so we briefly sketch such a scheme.
A diattenuation of the form of Eq. \eqref{eq:quantum diattenuation two input outputs} acting on the two modes of the TMSV leads to the transformation
\eq{
    &\ket{\psi_{\mathrm{TMSV}}}\bra{\psi_{\mathrm{TMSV}}}\to \hat{\varrho}=\sum_{m,n=0}^\infty \left(\frac{1-q}{q}\right)^m\left(\frac{1-r}{r}\right)^n\ket{\psi_{m,n}}\bra{\psi_{m,n}},\\
    &\qquad\ket{\psi_{m,n}}=\frac{1}{\cosh\zeta}\sum_{N\geq m,n}\left(-\eu^{\iu\varphi}\sqrt{qr}\tanh\zeta\right)^N\sqrt{\binom{N}{m}\binom{N}{n}}\ket{N-m,N-n}.
} 
To lowest order in the amount of loss, the QFIM for this evolved state is
\eq{
    \mathbf{Q}\left(q,r;\ket{\psi_{\mathrm{TMSV}}}\right)=\frac{H}{2}\begin{pmatrix}
        \frac{1}{1-q}+\mathcal{O}\left(1\right)&-\frac{3H}{2}-2+\mathcal{O}\left(1-q,1-r\right)\\
        -\frac{3H}{2}-2+\mathcal{O}\left(1-q,1-r\right)&\frac{1}{1-r}+\mathcal{O}\left(1\right)
    \end{pmatrix}.
} This evidently performs as well as a pair of Fock states, outperforming coherent states in the small-loss limit, and could readily be used for quantum-enhanced multiparameter diattenuation estimation by taking advantage of the polarization correlations generated by Type II spontaneous parametric downconversion.

\subsection{Combined Measurements}
It is not always possible to isolate individual Mueller- or Jones-matrix components for estimation; many realistic scenarios require the simultaneous estimation of, say, rotation and diattenuation parameters. As may be expected, there are tradeoffs in the precision with which each of these parameters may be simultaneously estimated due to different probe states being optimally sensitive for different parameters. 

\subsubsection{Phase and Loss}
The study of how to achieve quantum enhancements in the simultaneous estimation of two unrelated polarimetry parameters is much younger than its single-parameter counterpart. \textcite{Crowleyetal2014} first showed that one cannot achieve the ultimate quantum limit in measuring both a rotation angle and an attenuation parameter. Even though the QFIM for estimating these parameters is diagonal, the qCRB is not saturable, precluding an estimate with zero correlation between the two parameters.\footnote{It is still possible to approach the lower bound in some parameter regimes \cite{Albarellietal2019}.} Perhaps the most disappointing result is that the Heisenberg-scaling performance of phase estimation is overwhelmed by the shot-noise scaling performance of loss estimation in such a scenario for most nonzero values of loss, even when one does not desire to estimate the attenuation parameter \cite{DemkowiczDobrzanskietal2009}, due to
physical constraints imposed by the Kramers-Kronig relations \cite{Giananietal2021arxiv}; this makes quantum-enhanced polarimetry even more of a challenge. The main conclusion is that one has to determine the relative importance of the parameters being estimated in order to best optimize the tradeoffs in their estimation \cite{Crowleyetal2014}. This type of tradeoff is present in other quantum estimation tasks including the measurment of a phase and a phase diffusion parameter \cite{Vidrighinetal2014,Altorioetal2015,Szczykulskaetal2017,Rocciaetal2017}. Possible methods to circumvent these tradeoffs include having the parameters be correlated \cite{Birchalletal2020} and using postselection \cite{HoKondo2021}.

\subsubsection{Ellipsometry}
Until now we have not made any distinction between polarimetry and ellipsometry, as they both investigate polarization properties of the electric field. Ellipsometry typically focuses on determining a specific pair of polarization parameters, namely the ratio between two reflection or transmission components and the difference between two phase shifts, which can be cast as the ratio between two diattenuation parameters $\left(1-q\right)/\left(1-r\right)$ or $q/r$ and a rotation angle about the $\mathbf{e}_3$ axis.\footnote{Reflection ellipsometry often singles out two linear polarization components while transmission ellipsometry prefers circular components; the mathematics for both are identical and the physical scenarios can be interconverted using waveplates.} This has been studied from a few quantum perspectives \cite{Abouraddyetal2001,Abouraddyetal2002entangledellipsometry,Toussaintetal2002,Toussaintetal2004,Grahametal2006,Rudnickietal2020,WangAgarwal2021arxiv} that we review here. First, one can show that particular quantum states, such as entangled photon pairs, can be used for measuring these two parameters \cite{Abouraddyetal2001,Abouraddyetal2002entangledellipsometry,Toussaintetal2002,Toussaintetal2004,Grahametal2006,WangAgarwal2021arxiv}. The QFIM for this measurement has recently been calculated and shown to outperform that of coherent states, especially in the small-absorption regime \cite{WangAgarwal2021arxiv}, similar to other attenuation measurement results. Then, the ultimate quantum limit for such a measurement was studied by~\textcite{Rudnickietal2020}, where it was found that certain squeezed states allow for a simultaneous estimate of both ellipsometry parameters that approaches the Heisenberg limit. This is done by recasting the parameters as eigenvalues of an operator like $\hat{S}_3$ and an exponential-of-the-relative-phase operator
\eq{
    \hat{E}=\sum_N \hat{E}^{(N)},\qquad\hat{E}^{(N)}=\ket{N,0}\bra{0,N}+\sum_{m=0}^{N-1}\ket{m,N-m}\bra{m+1,N-m-1}
} that serves to shift eigenstates of $\hat{S}_3$ by one quantum of angular momentum \cite{LuisSanchezSoto1993} and finding a probe state that minimizes the uncertainty relation between these two operators. The optimal states in the limit of a large input energy are squeezed vacuum states; for finite energy, the probability distribution of the coefficients $\psi_m$ in Eq. \eqref{eq:pure state Nth layer in terms of psim} is the Fourier transform of a Mathieu function. Such optimal ellipsometric measurements have yet to be performed for large input energies $H$.

\subsubsection{Depolarization}
Finally, we mention that estimating the weights $\lambda_i$ from a general nondeterministic Mueller matrix stemming from Eq. \eqref{eq:Mueller from Jones with weights} can never be done with better than shot-noise scaling. This significant result is due to an argument given by \textcite{Jietal2008} relating quantum estimation to programmable quantum channels. Quantum-enhanced polarimetry must therefore be reserved for particular scenarios in which it is worth the cost of preparing special probe states and the ranges of parameters being estimated fall within the appropriate regions discussed above (pure rotations, small loss, etc.).

\subsection{Higher-Order Polarimetry: Measuring Quantum Polarization}
No discussion of quantum polarimetry is complete without describing measurements capable of discerning quantum polarization properties about which classical polarimetry is ignorant. 
A quantum decomposition following the classical form, as given in Eq. \eqref{eq:quantum decomposition into classical}, is not unique, with $\left(N+1\right)^2-4$ remaining degrees of freedom. Similarly,
a pure-state decomposition following the classical form
\eq{
    \ket{\psi^{(N)}}=\sqrt{p}\ket{\psi_{\mathrm{pol}}^{(N)}}+\eu^{\iu\varphi}\sqrt{1-p}\ket{\psi_{\mathrm{unpol}}^{(N)}},\quad \bra{\psi_{\mathrm{pol}}^{(N)}}\hat{S}_\mu \ket{\psi_{\mathrm{unpol}}^{(N)}}=0
    \label{eq:psiN classical decomp}
} has $2N-7$ remaining parameters after the degree and direction of polarization have been specified.
The quantum polarization properties can be arranged in many ways, all of which deal with the extra degrees of freedom present in quantum states beyond the four given by the Stokes parameters.

Quantum tomography can be inspired by classical polarimetry for determining all of the free parameters of  a quantum state. By assuming the polarization state to be an arbitrary state of $N$ photons that are not necessarily indistinguishable and thus not necessarily in a symmetric state, measuring the $4^N$ generalized Stokes parameters operators 
\eq{
    \hat{S}_{\mu \nu \cdots \xi}=\sigma_\mu\otimes\sigma_\nu\otimes\cdots\otimes \sigma_\xi
} will provide enough information to fully determine the state
 \cite{Jamesetal2001}, where we have used a first-quantized notation to delineate the operators acting on each photon as $2\times 2$ Pauli matrices. Linear combinations of these operators may be easier to measure for polarimetry \cite{Lingetal2006PRA}, but this does not improve the exponential scaling of the number of measurements required for full tomography. Many other such methods have been proposed using tools from maximum likelihood estimation \cite{Hradil1997}, Bayesian inference \cite{BlumeKohout2010}, and principal component analysis \cite{Lloydetal2014PCA}, as well as using mutually unbiased bases \cite{Lawrenceetal2002,Romeroetal2005,Klimovetal2008MUB,AdamsomSteinberg2010}.
 The number of measurements to be performed can be reduced using techniques from compressed sensing when the quantum state in question is close to being pure in terms of its matrix rank \cite{Grossetal2010,Crameretal2010,Liu2011,Shabanietal2011,Flammiaetal2012,Liuetal2012,Kalevetal2015,Baldwinetal2016,Steffensetal2017,Riofrioetal2017,Bouchardetal19compressedsensing,Gianani2020,GilLopezetal2021arxiv} and these measurements can be done adaptively \cite{Ahnetal2019adaptivecompressionexpt,Ahnetal2019adaptivecompressionnumerics} and have their completeness verified by machine-learning techniques \cite{Teoetal2021}. In practice, overcomplete measurement sets with redundant information tend to be much more useful than minimal measurement sets \cite{deBurghetal2008,Zhu2014}.

The SU(2) symmetry underlying quantum polarization dramatically reduces the number of parameters needed to be estimated for full tomography \cite{MankoManko1997,DArianoetal2003,Tothetal2010}. Inspecting Eq. \eqref{eq:pure state Nth layer in terms of psim}, for example, the number of free parameters in a pure $N$-photon state is reduced from $2^{N+1}-2$ to $2N$ and in a mixed state from $4^N-1$ to $\left(N+1\right)^2-1$. This makes the determination of all higher-order polarization properties feasible in practice.

How is quantum polarization best determined?
The minimal set of projection operators required for fully determining quantum polarization have been explicitly determined but are challenging to implement experimentally because they require entangled measurements \cite{Klimovetal2013}. 
In an analogous system with SU(2) symmetry, the basis states $\ket{m,N-m}$ can be spatially separated by applying an appropriate magnetic field as in a Stern-Gerlach experiment, facilitating a reconstruction of the quantum state using projectors \cite{DArianoetal2003}.
\eq{
    \hat{P}_m=\ket{m,N-m}\bra{m,N-m}
} 
Measurement of the projectors can be used to fully determine the properties of a mixed quantum polarization state if one measures these projectors as well as their rotated counterparts for an appropriate set of rotation angles
\cite{MankoManko1997}
\eq{
    \hat{P}_m\left(\btheta\right)=\hat{R}\left(\btheta\right)\ket{m,N-m}\bra{m,N-m}\hat{R}^\dagger\left(\btheta\right).
} 
This facilitates the experimental reconstruction of the SU(2) Wigner function that completely describes the quantum polarization state \cite{Mulleretal2012}.
Instead of spatially separating the components $\ket{m,N-m}$, they can be determined by the correlations between the intensities at the two detectors in the SU(2) gadget of Fig. \ref{fig:SU2 gadget}, through a scheme that can be traced back to \textcite{MukundaJordan1966}. First, using the SU(2) gadget with nine specific waveplate orientations allows one to directly measure all of the variances and covariances between the Stokes operators \cite{AgarwalChaturvedi2003}.
Next, if only the difference between the intensities at the two detectors is recorded for a sufficient variety of rotation angles, amounting to determining the expectation value of $\hat{P}_m(\btheta)$ summed over all $N\geq m$, a complete polarization tomogram can be constructed \cite{Bushevetal2001,KarassiovMasalov2004,Karassiov2005,Karassiov2007Pquasispin}. This can also be used to furnish an analysis of a cumulants for the Stokes parameters to arbitrary orders \cite{ChirkinSingh2021}, which is another way of arranging the higher-order polarization properties present in a quantums state.
Recording \textit{all} of the intensity correlations for the gadget in Fig. \ref{fig:SU2 gadget} amounts to measuring the expectation values of the correlation operators \cite{Schillingetal2010}
\eq{
    \hat{W}_m=\hat{a}^{\dagger m}\hat{b}^{\dagger N-m}\ha^m\hb^{N-m}
} 
that, when rotated appropriately, facilitate a complete reconstruction of the quantum polarization state \cite{Israeletal2012}. The intensity correlations can be directly determined using detectors with sufficient photon-number resolution \cite{Bayraktaretal2016}.
Alternatively, one can measure correlations between Stokes operators such as \cite{delaHozetal2013}
\eq{
    \hat{T}_l=\hat{S}_3^l
} and its rotated counterparts, where again the particular sets of sufficient rotations have been studied \cite{FilippovManko2010}.
These types of correlation measurements allow one to express the quantum polarization properties in a hierarchy of multipole moments \cite{Goldbergetal2021multipolesarxiv} that generalizes the according hierarchy in coherence theory \cite{Wolf2007}.

Another method for determining quantum polarization properties is measuring projectors onto SU(2)-coherent states
\eq{
    \hat{P}\left(\Omega\right)=\ket{\Omega}\bra{\Omega}
} and appropriately normalizing the results due to the overcompleteness of these states. This is possible due to the Husimi $Q$-function
\eq{
    q(\Omega)=\expct{\hat{P}\left(\Omega\right)}
} containing complete information about a quantum state \cite{Husimi1940}, which requires obtaining $q(\Omega)$ for a sufficient number of coordinates $\Omega$. For estimating polarization rotations, only four such measurements tend to be necessary \cite{Goldbergetal2021rotationspublished}, while the higher-order quantum polarization properties require $\left(N+1\right)^2-1$ measurements along a sufficient set of directions to completely specify the state \cite{HoffmannTakeuchi2004}.

The aforementioned compressed sensing techniques can even be used within the symmetric subspace to further enhance the efficiency of quantum polarization measurements for near-pure states \cite{Schwemmeretal2014}, with the current state of affairs reviewed by \textcite{TeoSanchezSoto2021accepted}.

We mention in conclusion that the higher-order polarization properties undergo higher-order transformations, leading to objects such as Mueller matrices with dimensions larger than $4\times 4$
\cite{Samimetal2016}. This type of polarimetry could benefit from quantum process tomography and may be useful to understanding high harmonic generation.

\section{Evaluating Classical Intuitions in Light of Quantum Polarimetry}
\label{sec:classical intuitions}
Armed with a full description of quantum polarization and its transformations, what do we have to say to classical physicists studying polarization? Different nuances are required in different scenarios, so we here highlight a few that we deem most relevant.

\subsection{Which Photons are Measured Classically?}
No detector measures every photon in an incoming beam of light. Given that the photonic nature underlying classical beams adds extra richness to their polarization properties, how does this affect the outcome of a classical polarization measurement? We here explain that the polarization properties of any subset of photons from a beam of light mimic the polarization properties of the whole beam from a classical standpoint.

We begin by inspecting a general quantum state with some fixed number of photons $N$:
\eq{
    \hat{\rho}^{(N)}=\sum_{m,n=0}^N\rho_{m,n}^{(N)}\ket{m,N-m}\bra{n,N-n}.
} Each basis state can be rewritten in a first-quantized description as a symmetric (i.e., permutationally invariant) superposition of single-photon states through
\eq{
    \ket{m,N-m}=\frac{1}{\sqrt{\binom{N}{m}}}\sum_{\mathrm{permutations}}\ket{\mathrm{R}}^{\otimes m}\otimes\ket{\mathrm{L}}^{\otimes N-m}.
} Then, if a detector only receives $N-1$ of the photons comprising the state, we can describe the state seen by the detector as the result of ignoring one of the permutationally invariant photons, which we set to be the final one for convenience:
\eq{
    \hat{\rho}^{(N-1)}=\Tr_{\mathrm{one\,photon}}\left(\hat{\rho}^{(N)}\right)=\sum_{\mathrm{X}=\mathrm{R},\mathrm{L}}\bra{X}_N\hat{\rho}^{(N)}\ket{X}_N.
} After performing the calculation for each basis state, we find the relationship between the state coefficients to be
\eq{
    \rho_{m,n}^{(N-1)}=\frac{\sqrt{\left(N-m\right)\left(N-n\right)}}{N}\rho_{m,n}^{(N)}+\frac{\sqrt{\left(m+1\right)\left(n+1\right)}}{N}\rho_{m+1,n+1}^{(N)},
} which may be iterated to find the coefficients for any $\hat{\rho}^{(M)}$ with $0\leq M\leq N$. This immediately yields interesting relationships, such as that obtaining any number of photons from a perfectly polarized or completely isotropic state with definite photon number yields a state with identical properties:
\eq{
    \Tr_{\mathrm{one\,photon}}\left(\ket{\Omega^{(N)}}\bra{\Omega^{(N)}}\right)=\ket{\Omega^{(N-1)}}\bra{\Omega^{(N-1)}},\qquad
    \Tr_{\mathrm{one\,photon}}\left(\hat{\mathds{1}}_N\right)=\hat{\mathds{1}}_{N-1}.
} Even more, we find that the Stokes parameters for the states remain the same up to the intensity normalization, with \cite{Goldberg2021thesis}
\eq{
    \frac{1}{N}\Tr\left(\rho_{m,n}^{(N)}\hat{S}_\mu\right)=\frac{1}{M}\Tr\left(\rho_{m,n}^{(M)}\hat{S}_\mu\right);
} the degree and direction of any subset $M$ photons from an $N$-photon beam are identical. This is a scenario in which quantum polarization reinforces classical intuition, even though the description of the light seems radically different from the latter. At the extreme of $M=1$, \textit{every photon that one examines from a beam with $N$ photons will reproduce the same polarization properties as the whole beam} [this was noted for $N=2$ by \textcite{Dograetal2020}].

We can ask the same question when given a beam of light with indeterminate photon number
\eq{
    \hat{\rho}=\sum_N \rho_N \hat{\rho}^{(N)}.
} Following immediately from the above results, we learn that, when $\hat{\rho}\to\Tr_{\mathrm{one\,photon}}\left(\hat{\rho}\right)$, the Stokes operators transform as
\eq{
    S_\mu=\sum_N\rho_N\Tr\left(\hat{\rho}^{(N)}\hat{S}_\mu\right)\to \sum_N\rho_N\frac{N-1}{N}\Tr\left(\hat{\rho}^{(N)}\hat{S}_\mu\right).
} When the Stokes parameters within each photon-number subspace have the same direction and degree of polarization, i.e., when
\eq{
    \Tr\left(\hat{\rho}^{(N)}\hat{S}_\mu\right)=Ns_\mu
} for some $N$-independent constants $s_\mu$, the overall direction and degree of polarization will be independent from the subset of photons that one obtains from the entire beam. If, however, the polarization properties are not the same in the different Fock layers, then inspecting $N$ photons will, in general, lead to different polarization properties than inspecting some other number $M$ photons. This forces us to distrust classical polarization properties whenever there is a possibility that different photon-number subspaces are polarized to different amounts or in different directions.

The stark differences between states of light with fixed and indeterminate numbers of photons emphasizes the additional complexities and richness that arise when quantum polarization properties are brought to bear.

\subsection{Rotating Unpolarized Light leads to Measurable Changes}
In classical optics, all unpolarized states of light are the same, up to specifying the total intensity of the light. Remarkably, this same conclusion does not follow quantum mechanically, even for the quantum states of light deemed to be closest to their classically described counterparts.

To recapitulate, we mentioned in Section \ref{sec:unpolarized states} that certain unpolarized quantum states of light are completely distinguishable from themselves after undergoing a polarization rotation. This is a marked difference from classically unpolarized light, as classical polarization dictates that unpolarized states do not transform when subject to polarization rotations. Similar discrepancies abound.

One method for obtaining classically unpolarized light is by incoherently mixing two perfectly polarized beams of light with equal intensities and orthogonal polarization components. Quantum mechanically, we similarly find that the state
\eq{
    \hat{\rho}=\frac{1}{2}\ket{\Omega^{(N)}}\bra{\Omega^{(N)}}+\frac{1}{2}\ket{\Omega_\perp^{(N)}}\bra{\Omega_\perp^{(N)}}
    \label{eq:mixture of two SU(2) coherent states}
} is classically unpolarized, where $\braket{\Omega^{(N)}}{\Omega_\perp^{(N)}}=0$, paralleling the discussion surrounding Eq. \eqref{eq:NOON}. However, this state is far from isotropic: the variances of the Stokes parameters are not all the same. Choosing, for example, $\Omega$ to point along the $\mathbf{e}_3$ axis, we find
\eq{
    \Var_{\hat{\rho}} \hat{S}_1=\Var_{\hat{\rho}} \hat{S}_2=\frac{N}{4},\quad \Var_{\hat{\rho}} \hat{S}_3=\frac{N^2}{4},
} which is responsible for the suboptimal precision of NOON states in metrology seen in Eq. \eqref{eq:NOON state MSE}. One might think that this anisotropy arises because the states being mixed in Eq. \eqref{eq:mixture of two SU(2) coherent states} are not fully classical. However, a similar result arises when two canonical coherent states with orthogonal polarization properties are incoherently mixed:
\eq{
    \hat{\rho}=\frac{1}{2}\ket{\alpha_\Omega}\bra{\alpha_\Omega}+\frac{1}{2}\ket{\alpha_{\Omega_\perp}}\bra{\alpha_{\Omega_\perp}}.
} Again choosing $\Omega$ to point along the $\pmb{e}_3$ axis leads to the anisotropic Stokes parameter variances
\eq{
    \Var_{\hat{\rho}} \hat{S}_1=\Var_{\hat{\rho}} \hat{S}_2=\frac{\left|\alpha\right|^2}{4},\quad \Var_{\hat{\rho}} \hat{S}_3=\frac{\left|\alpha\right|^2\left(1+\left|\alpha\right|^2\right)}{4}.
} These anisotropies mean that \textit{the variance of a Stokes parameter will change if such a classically unpolarized state has its polarization rotated} and, further, that \textit{mixing a right- with a left-handed circularly polarized beam leads to a different state than mixing a horizontally with a vertically polarized beam}. Such information is completely lost in a classical description of polarization and yet these properties are completely discernable in classical polarization experiments. The only method to circumvent such problems is to acquire completely isotropic states of the form of Eq. \eqref{eq:isotropic state}, which may be obtained by the complete depolarization of a beam of light
\eq{
    \hat{\rho}\to\int d\btheta\, \hat{R}\left(\btheta\right)\hat{\rho}\hat{R}\left(\btheta\right)^\dagger.
} These discrepancies can all be attributed to the plethora of quantum states underlying a single classical decomposition of the form of Eq. \eqref{eq:coherency decomposition}.

\subsection{Transformations of the Polarized-Plus-Unpolarized Decomposition}
Polarization transformations act linearly on the polarized and unpolarized components in Eqs. \eqref{eq:Stokes decomposition} and \eqref{eq:coherency decomposition}. This means that we can inspect the action of all classical polarization transformations on polarized and unpolarized states \textit{in vacuo} and combine the actions for partially polarized states. We will see here that such a correspondence only works for the quantum decomposition of Eq. \eqref{eq:quantum decomposition into classical}, even though other decompositions have other strong physical motivations.

We will iterate three possible quantum decompositions underlying the classical one, following Eqs. \eqref{eq:quantum decomposition into classical}, \eqref{eq:psiN classical decomp}, and more:
\eq{
    &\mathrm{I:}\,\,\,\ket{\psi}=\sqrt{p}\ket{\psi_{\mathrm{pol}}}+\eu^{\iu\varphi}\sqrt{p}\ket{\psi_{\mathrm{unpol}}},\quad \bra{\psi_{\mathrm{unpol}}}\hat{S}_\mu\ket{_{\mathrm{pol}}}=0,\\
    &\mathrm{II:}\,\,\,\hat{\rho}=p\hat{\rho}_{\mathrm{pol}}+\left(1-p\right)\hat{\rho}_{\mathrm{unpol}},\\
    &\mathrm{III:}\,\hat{\rho}=p\hat{\rho}_{\mathrm{pol}}+\left(1-p\right)\hat{\rho}_{\mathrm{isotropic}}.
} Each of these decompositions transforms as expected classically under polarization rotations, in that the polarized component remains polarized in a new direction as per the classical rotation of the Stokes vector, the unpolarized component remains unpolarized, and the degree of polarization remains unchanged. Such agreement will not be seen with other polarization transformations. We can also inspect the minimum number of nonclassical polarization degrees of freedom in each decomposition, which happens in a single $N$-photon subspace, where I, II, and III have $2N-7$,  $\left(N+1\right)^2-4$, and $0$ remaining degrees of freedom. This points to III best resembling the classical decomposition, but we will see this to not be the case following polarization transformations. Of note, pure, partially polarized states such as $\ket{m,n}$ with $m\neq n$ and $m,n\neq 0$ obey none of the decompositions, so the classical decomposition cannot be taken for granted.

Diattenuation and depolarization transformations, in general, reduce the purity of an input quantum state. This means that the pure-state decomposition I will no longer hold following a general polarization transformation, so we cannot simply say that quantum superpositions underly the classical decompositions into polarized and unpolarized fractions.

Partial attenuation 
ruins decomposition III. Choosing $q=1$ and a nontrivial value of $r$ for the attenuation of the second mode, an isotropic unpolarized state transforms into
\eq{
    \frac{\hat{\mathds{1}}_N}{N+1}\quad\underset{r}{\to}\quad\frac{1}{N+1}\sum_{m=0}^N\sum_{M=m}^{N}\binom{N-m}{M-m}r^{M-m}\left(1-r\right)^{N-M}\ket{m,M-m}\bra{m,M-m}.
} The unpolarized component of this state is only isotropic in the vacuum and $M=1$ photon-number subspaces; otherwise, it no longer follows decomposition III. This is why we generally state that only decomposition II, i.e., Eq. \eqref{eq:quantum decomposition into classical}, may underlie the classical decomposition, even though decomposition III is preferable in terms of degrees of freedom and isotropy properties.

It is clear from the discussion of unpolarized states that decomposition III will also perform poorly for depolarization transformations. For example, we expect a convex combination of a do-nothing operation and rotation by $\pi$ to be able to completely depolarize an incident polarized state:
\eq{
    \hat{\rho}_{\mathrm{pol}}\quad\underset{\mathrm{depolarization}}{\to}\quad\frac{1}{2}\hat{\rho}_{\mathrm{pol}}+\frac{1}{2}\hat{R}\left(\pi,\mathbf{n}\right)\hat{\rho}_{\mathrm{pol}}\hat{R}^\dagger\left(\pi,\mathbf{n}\right)=\hat{\rho}_{\mathrm{unpol}}\neq\hat{\rho}_{\mathrm{isotropic}},
} where $\mathbf{n}$ may be any axis orthogonal to the direction of polarization of $\hat{\rho}_{\mathrm{pol}}$. Other depolarizing transformations do indeed suffice for decomposition III, such as that of complete depolarization
\eq{
    \hat{\rho}_{\mathrm{pol}}\quad\underset{\mathrm{complete\,depolarization}}{\to}\quad\int d\btheta=\hat{R}\left(\btheta\right)\hat{\rho}_{\mathrm{pol}}\hat{R}^\dagger\left(\btheta\right)=\hat{\rho}_{\mathrm{unpol}}=\hat{\rho}_{\mathrm{isotropic}}.
} 

From these considerations, we see that quantum decompositions I-III may be seen to underly the classical polarization decomposition in certain circumstances, but only the quantum decomposition into first-order polarized and unpolarized fractions, II, transforms correctly under all classical polarization transformations. We are thus faced with two alternatives: \textit{give up the privileged status of classical polarization transformations} or \textit{give up the physical intuition of unpolarized states being isotropic under polarization rotations}. Both alternatives are required for the quantum states that have no analogue to the classical decomposition into polarized and unpolarized fractions.

\subsection{Higher-Order Correspondences}
Quantum polarization dictates that classical polarization properties arise from expectation values of noncommuting operators. Therefore, \textit{we should not expect higher-order moments of Stokes parameters to transform classically}.
The Stokes operators themselves transform classically under rotations, so we will leave those aside in this section.

We elucidate the present discrepancy with the example of a diattenuation along the $\mathbf{e}_3$ axis, which leads to the classical and quantum transformation
\eq{
    S_3\quad\underset{q,r}{\to}\quad\frac{q-r}{2}S_0+\frac{q+r}{2}S_3.
} Squared, we expect a transformation like
$S_3^2\to\left(\frac{q-r}{2}S_0+\frac{q+r}{2}S_3\right)^2$, which is quadratic in the Stokes parameters. Quantum mechanically, however, we must take into account the noncommutativity of the vacuum operators $\left[\hat{v}_i,\hat{v}_j^\dagger\right]=\delta_{i,j}$ before tracing out their modes, leading to
\eq{
    \hat{S}_3^2\quad\underset{q,r}{\to}\quad\left(\frac{q-r}{2}\hat{S}_0+\frac{q+r}{2}\hat{S}_3\right)^2+\frac{q\left(1-q\right)}{4}\left(\hat{S}_0+\hat{S}_3\right)+\frac{r\left(1-r\right)}{4}\left(\hat{S}_0-\hat{S}_3\right).
} This differs from what is expected classically, even though the operators $\hat{S}_0$ and $\hat{S}_3$ commute. In the limit of strong fields, the terms linear in the Stokes operators may perhaps be neglected and one may be permitted to approximate $\expct{\hat{S}_0\hat{S}_3}\approx S_0 S_3$, but a careful consideration of the diattenuation parameters must be considered in such a case to ensure that the fields remain sufficiently strong.

The deviations from classical predictions are even more pronounced with depolarization transformations of higher-order Stokes parameters. 
This is because a product of Stokes operators evolving under a Kraus map of the form of Eq. \eqref{eq:quantum channel Kraus on Stokes} undergoes
\eq{
    \hat{S}_\mu\hat{S}_\nu\to\sum_l\hat{K}_l^\dagger \hat{S}_\mu\hat{S}_\nu\hat{K}_l=\sum_{l,m}\hat{K}_l^\dagger \hat{S}_\mu\hat{K}_m^\dagger \hat{K}_m\hat{S}_\nu\hat{K}_l,
} where the operations $\hat{K}_m\hat{S}_\nu\hat{K}_l$ are never guaranteed to equal linear combinations of Stokes operators. In some cases, they do facilitate a decomposition into such linear combinations, but those linear combinations are not the ones predicted classically. 
One must therefore be leery of all classical polarization transformations calculated with higher-order Stokes parameters and use a quantum formalism for understanding any such process.

\section{Conclusions}
We have toured quantum polarization and the impacts it has on polarimetry. Along the way, we saw the impacts of quantum polarimetry on its classical counterpart and the possibilities of performing quantum-enhanced polarimetry and quantum polarimetry of manifestly nonclassical properties of light.
Light's polarization continues to be a subject of prime interest to novices and practitioners alike and we suspect more mysteries and features of the theory to be unearthed as we go deeper into the quantum realm.


\textit{Acknowledgments:} AZG would like to thank Girish Agarwal, Jos\'e Gil, Khabat Heshami, Daniel James, and Wenlei Zhang for discussions and Luis S\'anchez-Soto for insightful comments on the manuscript.

%

\end{document}